\documentclass[twocolumn]{aastex631}
\usepackage{graphicx}
\usepackage[normalem]{ulem}
\usepackage{float}
\usepackage[caption=false]{subfig}
\usepackage{enumitem}
\useunder{\uline}{\ul}{}
\usepackage{csquotes}
\usepackage{hyperref}

\newcommand\JWST{\textit{JWST }}

\newcommand\jwst{\texttt{jwst}} 

\begin{document}
\title{JADES Initial Data Release for the Hubble Ultra Deep Field: Revealing the Faint Infrared Sky with Deep \JWST NIRCam Imaging}

\author[0000-0002-7893-6170]{Marcia J.\ Rieke}
\affiliation{Steward Observatory, University of Arizona, 933 N. Cherry Ave,
Tucson, AZ 85721, USA}

\author[0000-0002-4271-0364]{Brant Robertson}
\affiliation{Department of Astronomy and Astrophysics, University of California, Santa Cruz, 1156 High Street, Santa Cruz, CA 95064, USA}

\author[0000-0002-8224-4505]{Sandro Tacchella}
\affiliation{Cavendish Laboratory, University of Cambridge, 19 JJ Thomson Avenue, Cambridge, CB3 OHE, UK}
\affiliation{Kavli Institute for Cosmology, Madingley Road, Cambridge, CB3 0HA, UK}

\author[0000-0003-4565-8239]{Kevin Hainline}
\affiliation{Steward Observatory, University of Arizona, 933 N. Cherry Ave, Tucson, AZ 85721, USA}

\author[0000-0002-9280-7594]{Benjamin D.\ Johnson}
\affiliation{Center for Astrophysics | Harvard and Smithsonian, 60 Garden Street, Cambridge, MA 02138, USA}

\author[0000-0002-8543-761X]{Ryan Hausen}
\affiliation{Department of Physics and Astronomy, The Johns Hopkins University, 3400 N. Charles St., Baltimore, MD 21218, USA}

\author[0000-0001-7673-2257]{Zhiyuan Ji}
\affiliation{Steward Observatory, University of Arizona, 933 N. Cherry Ave,
Tucson, AZ 85721, USA}

\author[0000-0001-9262-9997]{Christopher N.\ A.\ Willmer}
\affiliation{Steward Observatory, University of Arizona, 933 N. Cherry Ave,
Tucson, AZ 85721, USA}

\author[0000-0002-2929-3121]{Daniel J.\ Eisenstein}
\affiliation{Center for Astrophysics | Harvard and Smithsonian, 60 Garden Street, Cambridge, MA 02138, USA}

\author[0000-0001-8630-2031]{Dávid Puskás}
\affiliation{Cavendish Laboratory, University of Cambridge, 19 JJ Thomson Avenue, Cambridge, CB3 OHE, UK}
\affiliation{Kavli Institute for Cosmology, Madingley Road, Cambridge, CB3 0HA, UK}

\author[0000-0002-8909-8782] {Stacey Alberts}
\affiliation{Steward Observatory, University of Arizona, 933 N. Cherry Ave,
Tucson, AZ 85721, USA}

\author[0000-0001-7997-1640]{Santiago Arribas}
\affiliation{Centro de Astrobiología (CAB), CSIC–
INTA, Cra. de Ajalvir Km.~4, 28850- Torrejón de Ardoz, Madrid, Spain}

\author[0000-0003-0215-1104]{William M.\ Baker}
\affiliation{Cavendish Laboratory, University of Cambridge, 19 JJ Thomson Avenue, Cambridge, CB3 OHE, UK}
\affiliation{Kavli Institute for Cosmology, Madingley Road, Cambridge, CB3 0HA, UK}

\author[0000-0002-4735-8224]{Stefi Baum} 
\affiliation{Department of Physics and Astronomy, University of Manitoba, Winnipeg, MB R3T 2N2, Canada}

\author[0000-0003-0883-2226]{Rachana Bhatawdekar}
\affiliation{European Space Agency, ESA/ESTEC, Keplerlaan 1,
2201 AZ Noordwijk, NL}

\author[0000-0001-8470-7094]{Nina Bonaventura}
\affiliation{Steward Observatory, University of Arizona, 933 N. Cherry Ave,
Tucson, AZ 85721, USA}

\author[0000-0003-4109-304X]{Kristan Boyett}
\affiliation{School of Physics, University of Melbourne, Parkville 3010, VIC, Australia}
\affiliation{ARC Centre of Excellence for All Sky
Astrophysics in 3 Dimensions (ASTRO 3D), Australia}

\author[0000-0002-8651-9879]{Andrew J.\ Bunker}
\affiliation{Department of Physics, University of Oxford, Denys Wilkinson Building, Keble Road, Oxford OX1 3RH, UK}

\author[0000-0002-0450-7306]{Alex J.\ Cameron}
\affiliation{Department of Physics, University of Oxford, Denys Wilkinson Building, Keble Road, Oxford OX1 3RH, UK}

\author[0000-0002-6719-380X]{Stefano Carniani}
\affiliation{Scuola Normale Superiore, Piazza dei Cavalieri 7, I-56126 Pisa, Italy}

\author[0000-0003-3458-2275]{Stephane Charlot}
\affiliation{Sorbonne Université, UPMC-CNRS, UMR7095, Institut d'Astrophysique de Paris, F-75014 Paris, France}

\author[0000-0002-7636-0534]{Jacopo Chevallard}
\affiliation{Department of Physics, University of Oxford, Denys Wilkinson Building, Keble Road, Oxford OX1 3RH, UK}

\author[0000-0002-2178-5471]{Zuyi Chen}
\affiliation{Steward Observatory, University of Arizona, 933 N. Cherry Ave,
Tucson, AZ 85721, USA}

\author[0000-0002-2678-2560]{Mirko Curti}
\affiliation{Cavendish Laboratory, University of Cambridge, 19 JJ Thomson Avenue, Cambridge, CB3 OHE, UK}
\affiliation{Kavli Institute for Cosmology, Madingley Road, Cambridge, CB3 0HA, UK}

\author[0000-0002-9551-0534]{Emma Curtis-Lake }
\affiliation{Centre for Astrophysics Research, Department
of Physics, Astronomy and Mathematics, University of Hertfordshire, Hatfield AL10 9AB, UK}

\author[0000-0002-9708-9958]{A.\ Lola Danhaive}
\affiliation{Kavli Institute for Cosmology, Madingley Road, Cambridge, CB3 0HA, UK}

\author[0000-0002-4781-9078]{Christa DeCoursey}
\affiliation{Steward Observatory, University of Arizona, 933 N. Cherry Ave,
Tucson, AZ 85721, USA}

\author[0000-0002-6317-0037]{Alan Dressler} 
\affiliation{The Observatories of the Carnegie Institution for Science, 813 Santa Barbara St., Pasadena, CA 91101}

\author[0000-0003-1344-9475]{Eiichi Egami}
\affiliation{Steward Observatory, University of Arizona, 933 N. Cherry Ave,
Tucson, AZ 85721, USA}

\author[0000-0003-4564-2771]{Ryan Endsley}
\affiliation{Department of Astronomy, University of Texas, Austin, TX 78712 USA}

\author[0000-0003-4337-6211]{Jakob M.\ Helton }
\affiliation{Steward Observatory, University of Arizona, 933 N. Cherry Ave,
Tucson, AZ 85721, USA}

\author[0000-0002-4684-9005] {Raphael E.\ Hviding}
\affiliation{Steward Observatory, University of Arizona, 933 N. Cherry Ave, Tucson, AZ 85721, USA}

\author[0000-0002-5320-2568]{Nimisha Kumari}
\affiliation{AURA for European Space Agency, Space Telescope Science
Institute, 3700 San Martin Drive, Baltimore, MD 21218, USA.}

\author[0000-0002-3642-2446]{Tobias J.\ Looser}
\affiliation{Cavendish Laboratory, University of Cambridge, 19 JJ Thomson Avenue, Cambridge, CB3 OHE, UK}
\affiliation{Kavli Institute for Cosmology, Madingley Road, Cambridge, CB3 0HA, UK}

\author[0000-0002-6221-1829]{Jianwei Lyu }
\affiliation{Steward Observatory, University of Arizona, 933 N. Cherry Ave,
Tucson, AZ 85721, USA}

\author[0000-0002-4985-3819]{Roberto Maiolino} 
\affiliation{Cavendish Laboratory, University of Cambridge, 19 JJ Thomson Avenue, Cambridge, CB3 OHE, UK}
\affiliation{Kavli Institute for Cosmology, Madingley Road, Cambridge, CB3 0HA, UK}
\affiliation{Department of Physics and Astronomy, University College London, Gower Street, London
WC1E 6BT, UK}

\author[0000-0003-0695-4414]{Michael V.\ Maseda }
\affiliation{Department of Astronomy, University of Wisconsin-Madison, 475 N. Charter St., Madison, WI 53706 USA}

\author[0000-0002-7524-374X]{Erica J.\ Nelson}
\affiliation{Department for Astrophysical and Planetary Science, University of Colorado, Boulder, CO 80309, USA}

\author[0000-0003-2303-6519]{George Rieke}
\affiliation{Steward Observatory, University of Arizona, 933 N. Cherry Ave,
Tucson, AZ 85721, USA}

\author[0000-0003-4996-9069]{Hans-Walter Rix}
\affiliation{Max-Planck-Institut für Astronomie, Königstuhl 17, D-69117, Heidelberg, Germany}
 
\author[0000-0001-9276-7062]{Lester Sandles}
\affiliation{Cavendish Laboratory, University of Cambridge, 19 JJ Thomson Avenue, Cambridge, CB3 OHE, UK}
\affiliation{Kavli Institute for Cosmology, Madingley Road, Cambridge, CB3 0HA, UK}

\author[0000-0001-5333-9970]{Aayush Saxena}
\affiliation{Department of Physics and Astronomy, University College London, Gower Street, London WC1E 6BT, UK}

\author[0000-0001-8225-8969]{Katherine Sharpe} 
\affiliation{Center for Astrophysics $|$ Harvard \& Smithsonian, 60 Garden St., Cambridge MA 02138 USA}

\author[0000-0003-4702-7561]{Irene Shivaei}
\affiliation{Steward Observatory, University of Arizona, 933 N. Cherry Ave,
Tucson, AZ 85721, USA}
\affiliation{Centro de Astrobiología (CAB), CSIC-INTA, Ctra. de Ajalvir km 4, 
Torrejón de Ardoz, E-28850, Madrid, Spain}

\author[0009-0004-0844-0657]{Maya Skarbinski} 
\affiliation{Center for Astrophysics $|$ Harvard \& Smithsonian, 60 Garden St., Cambridge MA 02138 USA}

\author[0000-0001-8034-7802]{Renske Smit}
\affiliation{Astrophysics Research Institute, Liverpool John Moores University,
146 Brownlow Hill, Liverpool L3 5RF, UK}

\author[0000-0001-6106-5172]{Daniel P.\ Stark}
\affiliation{Steward Observatory, University of Arizona, 933 N. Cherry Ave,
Tucson, AZ 85721, USA}

\author[0000-0002-9720-3255]{Meredith Stone} 
\affiliation{Steward Observatory, University of Arizona, 933 N. Cherry Ave,
Tucson, AZ 85721, USA}

\author[0000-0002-1714-1905]{Katherine A. Suess}
\affiliation{Department of Astronomy and Astrophysics, University of California, Santa Cruz, Santa Cruz, CA 95064, USA}
\affiliation{Kavli Institute for Particle Astrophysics and Cosmology and Department of Physics, Stanford University, Stanford, CA 94305, USA}

\author[0000-0002-4622-6617]{Fengwu Sun}
\affiliation{Steward Observatory, University of Arizona, 933 N. Cherry Ave,
Tucson, AZ 85721, USA}

\author[0000-0001-8426-1141]{Michael Topping}
\affiliation{Steward Observatory, University of Arizona, 933 N. Cherry Ave,
Tucson, AZ 85721, USA}

\author[0000-0003-4891-0794]{Hannah \"Ubler }
\affiliation{Cavendish Laboratory, University of Cambridge, 19 JJ Thomson Avenue, Cambridge, CB3 OHE, UK}
\affiliation{Kavli Institute for Cosmology, Madingley Road, Cambridge, CB3 0HA, UK}

\author[0000-0001-6917-4656]{Natalia C.\ Villanueva} 
\affiliation{Center for Astrophysics $|$ Harvard \& Smithsonian, 60 Garden St., Cambridge MA 02138 USA}

\author[0000-0002-0695-8485]{Imaan E.\ B.\ Wallace} 
\affiliation{Department of Physics, University of Oxford, Denys Wilkinson Building, Keble Road, Oxford OX1 3RH, UK}

\author[0000-0003-2919-7495]{Christina C. Williams}
\affiliation{NSF’s National Optical-Infrared Astronomy Research Laboratory, 950 North Cherry Avenue, Tucson, AZ 85719, USA}

\author[0000-0002-4201-7367]{Chris Willott}
\affiliation{NRC Herzberg, 5071 West Saanich Rd, Victoria, BC V9E 2E7, Canada}

\author[0000-0003-1432-7744] {Lily Whitler}
\affiliation{Steward Observatory, University of Arizona, 933 N. Cherry Ave, Tucson, AZ 85721, USA}

\author[0000-0002-7595-121X] {Joris Witstok}
\affiliation{Cavendish Laboratory, University of Cambridge, 19 JJ Thomson Avenue, Cambridge, CB3 OHE, UK}
\affiliation{Kavli Institute for Cosmology, Madingley Road, Cambridge, CB3 0HA, UK}

\author[0000-0001-5962-7260]{Charity Woodrum} 
\affiliation{Steward Observatory, University of Arizona, 933 N. Cherry Ave,
Tucson, AZ 85721, USA}

\correspondingauthor{Marcia J. Rieke}
\email{mrieke@arizona.edu}

\begin{abstract}
JWST  has revolutionized the field of extragalactic astronomy with its sensitive and high-resolution infrared view of the distant universe. Adding to the new legacy of JWST observations, we present the first NIRCam imaging data release from the JWST Advanced Deep Extragalactic Survey (JADES) providing 9 filters of infrared imaging of $\sim$25 arcmin$^2$ covering the Hubble Ultra Deep Field and portions of Great Observatories Origins Deep Survey (GOODS) South.  Utilizing 87 on-sky dual-filter hours of exposure time, these images reveal the deepest ever near-infrared view of this iconic field.  We supply carefully constructed 9-band mosaics of the JADES bands, as well as matching reductions of 5 additional bands from the JWST Extragalactic Medium-band Survey (JEMS).  Combining with existing HST imaging, we provide 23-band space-based photometric catalogs and photometric redshifts for $\approx47,500$ sources.
To promote broad engagement with the JADES survey, we have created an interactive {\tt FitsMap} website 
to provide an interface for professional researchers and the public
to experience these JWST datasets.
Combined with the first JADES NIRSpec data release, these public JADES imaging and spectroscopic
datasets provide a new foundation for discoveries of the infrared universe by the worldwide
scientific community.
\end{abstract}

\section{Introduction}
An essential route to the study of the early history of galaxies is the
detection and characterization of incredibly distant galaxies,
utilizing the finite light travel time to see galaxies in their youth.
This has driven astronomers to build and use telescopes to study
extremely faint and small galaxies.  The James Webb Space Telescope \citep{Gardner_etal2023}
is the next great step, optimized for the essential wavelengths in the
infrared and equipped with extremely flexible and capable instruments.
The first year of the mission has revealed high-redshift galaxies
as never before possible (\citep{robertson_etal23, curtislake2023}

In the past three decades, astronomers have focused increasingly on
deep blank-field surveys, as these give an unbiased selection of high-redshift
galaxies, the populations of which reveal themselves at the faintest
technological limits.  A famous step in this quest was the Hubble
Deep Field \citep[HDF;][]{williams1996a,ferguson2000a}, which revealed a dazzling
array of galaxies beyond redshift $z\sim3$.  This prompted heavy investment
from many large telescopes in this field as well as in a southern
companion, the Chandra Deep Field South \citep[CDF-S;][]{giacconi2002a}, and
both fields were expanded in the Great Observatories Origins Deep
Survey \citep[GOODS;][]{giavalisco2004a}.  Soon after, the Hubble Ultra
Deep Field (HUDF), the deepest optical image yet taken, was observed at the
center of the GOODS-S field \citep{beckwith2006a}.

NICMOS was used for the first space-based infrared observations of the HDF \citep{thompson1999}.
Other studies such as \cite{conselice2011} exploited the capabilities provided by 
NICMOS to study the evolution of massive galaxies. The installation of Wide Field Camera 3 on
HST afforded improved infrared sensitivity and a substantially wider field of view which led
to much deeper 1 to 2 micron imaging (eg., Bouwens et al. \citeyear{bouwens2010},
Oesch et al. \citeyear{oesch2010}). This improved infrared
imaging provided much larger z$\sim$7 samples than previously available. 

The two deep fields GOODS-S and GOODS-N, as well as a small number of others, have attracted
enormous collective attention from essentially every large narrow-field
telescope in the world, as astronomers interested in the faint extragalactic
sky pour their new investments of observing time into the fields to take
advantage of the previous investments.  The HUDF has remained the
premier field, with contributions too numerous to list but hosting
the deepest general-purpose extragalactic images at nearly all common
wavelengths.  Key examples include very deep HST infrared images,
such as from the HUDF09 and HUDF12 program \citep{ellis2013a,illingworth2013a},
the continued investment of Chandra imaging \citep[e.g.,][]{xue2016a},
deep ALMA observations \citep{aravena2016a,walter2016a,dunlop2017a, hatsukade18}, 
deep JVLA imaging \citep[e.g.,][]{rujopakarn2016a},
and large investments of spectroscopy, such as with VLT MUSE \citep[e.g.,][]{bacon2017a,bacon2021a}.

With the JWST Advanced Deep Extragalactic Survey (JADES), we aim to continue the
legacy of the GOODS-S/HUDF and GOODS-N/HDF fields, bringing extremely
deep high-quality JWST near-infrared imaging and spectroscopy to the
field.  JADES is a collaboration of the JWST Near-Infrared Camera
(NIRCam) and Near-Infrared Spectrograph instrument development teams, pooling
about 770 hours of guaranteed time of the mission to the purpose
of executing a carefully crafted survey of the fields \citep{eisenstein23r}.

Here, we present the initial data release of NIRCam imaging from JADES,
covering 25 arcmin$^2$ on and around the HUDF.  This 9-band imaging includes
87 open-shutter hours of dual-filter imaging, comprising about 111 hours
of mission time.  Exposure times range from $\sim$14 to $\sim$60~ks per
filter.  Our release mosaics and catalogs include 5 further bands
from the JWST Extragalactic Medium-band Survey \citep[JEMS][]{williams2023a},
which covers one-third of the release area.  A companion paper \citep{bunker23r} presents deep spectroscopy on the HUDF.

We describe the data included in this release in \S~\ref{sec:contents}, the image processing methodology in \S~\ref{sec:mosaics}, and the photometry methods in \S~\ref{sec:catalog}.  In \S~\ref{sec:data_properties}, we summarize the quantitative performance assessment and validation of the 
images and catalogs, although further information will be presented in Tacchella et al.\ (in prep.) and Robertson et al.\ (in prep.). Our
\S~\ref{sec:photoz} presents a set of photometric redshifts computed from the catalog, utilizing up to 14 JWST bands and 5 HST ACS bands, with full details provided by \citep{hainline23}.  We conclude in \S~\ref{sec:conclusions}.

\begin{figure}[ht!]
\centering
\includegraphics[width=8.5cm]{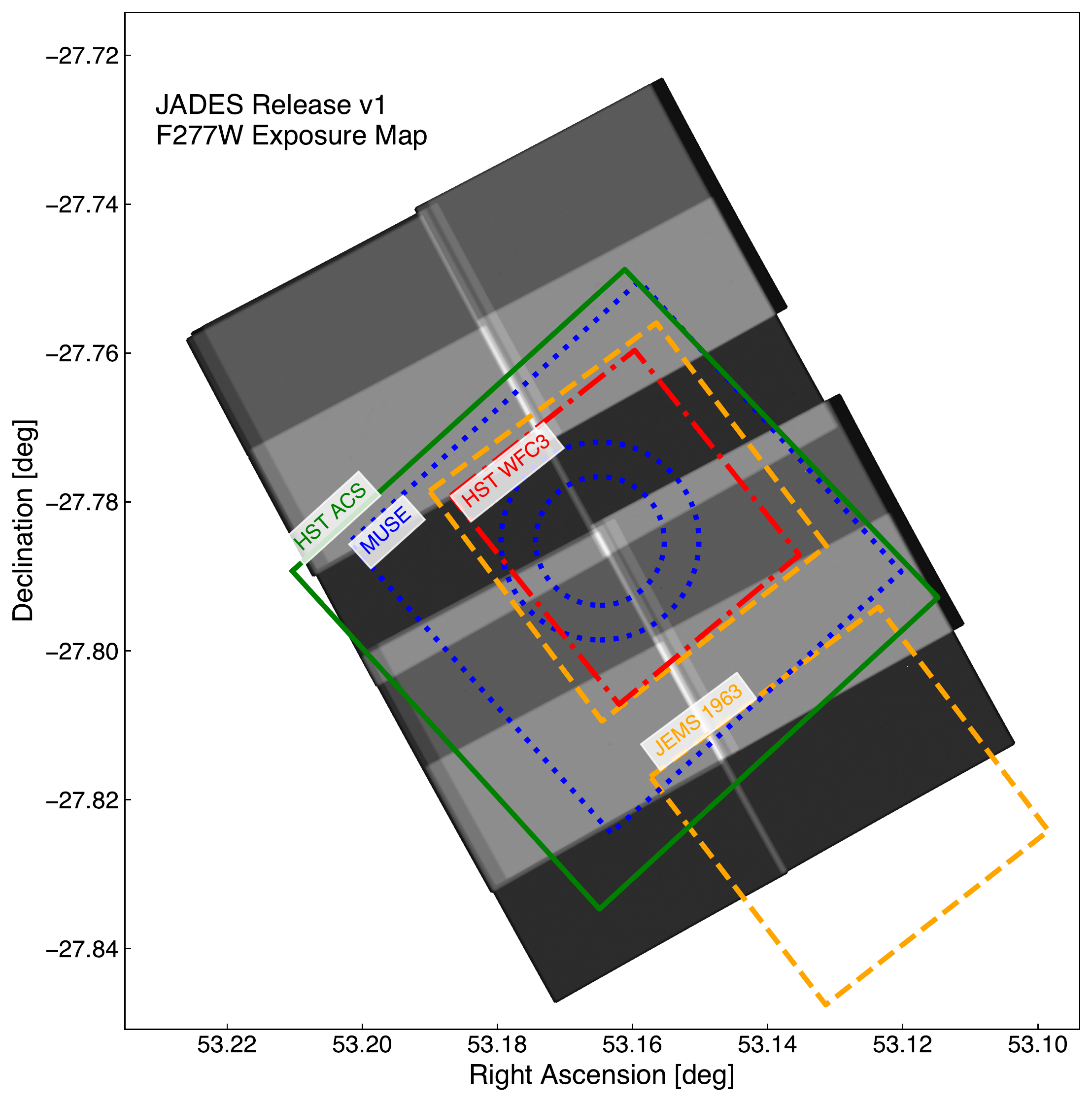}
\caption{Field layout of the first JADES data release images relative to other surveys in the GOODS-S region.
The coverage of the JADES F277W imaging in this first data release is shown as a grayscale image, and is representative of the coverage in each JADES filter. This portion of the JADES survey covers the HST WFC3 Ultra Deep Field \citep[red dash-dot;][]{ellis2013a,illingworth2013a}, the HST ACS Ultra Deep Field \citep[green solid;][]{beckwith2006a}, the MUSE Ultra Deep Field mosaic \citep[blue, dotted square][]{bacon2017a}, and the MUSE Extremely Deep Field \citep[blue, dotted circle][]{bacon2021a}. The blue circles outline the 10-hour and 100-hour MUSE exposure depths. The JWST Extragalactic Medium-band Survey \citep[JEMS;][]{williams2023a}, shown using an orange,dashed line, overlaps with the HST UDF and the JADES NIRCam imaging. We incorporate the JEMS imaging into this release, with uniform processing.
}
\label{fig:release_area}
\end{figure}

\section{Data Release Contents}\label{sec:contents}
The imaging portion of this first JADES data release comes from the NIRCam imaging of program 1180 (PI: Eisenstein), Observations 7, 10, 11, 15, 17, and 18.  These are half of the Deep Prime mosaic, covering the Hubble Ultra Deep Field and a log of the observations is presented in Table~\ref{tab:log}, which tabulates several parameters used in the \textsl{Astronomer's Proposal Tool (APT)}, used to identify the observation, namely, the pointing name, the observation number, the name associated to pointing (TARGPROP), the visit identification, the starting and ending UT date of the observation. As described in \citet{eisenstein23r}, observations 7, 10, 15, and 18 form a 2 by 2 overlapping mosaic of pointings, each a 9-point dither of 5 filter pairs.  Observations 11 and 17 add another 8-point dither, building depth in 3 of the filter pairs in two of the pointings.  Each exposure is 1375 seconds.  These data were acquired from September 29 through October 5, 2022, at which time the zodiacal light background was low.  

\begin{deluxetable*}{lclccc}[htb]
\centering 
\tablecaption{Log of NIRCam Observations\label{tab:log}}
\tablewidth{0pt}
\tablehead{
\colhead{Pointing} & \colhead{Observation} &\colhead{TARGPROP~} &
\colhead{VISIT\_ID} & \colhead{Date start} &\colhead{Date end}\\
}
\startdata
Deep Pointing 1 Part 1 & 7 & POINTINGONE-B & 01180007001 & 2022-09-29 & 2022-09-30\\
Deep Pointing 2 Part 1 &10 & POINTINGTWO-B & 01180010001 & 2022-09-30 & 2022-10-01\\
Deep Pointing 2 Part 2 &11 & POINTINGTWO-C & 01180011001 & 2022-10-03 & 2022-10-04\\
Deep Pointing 3 Part 3 &15 & POINTINGTHREE-A & 01180015001 &2022-10-04 & 2022-10-05\\
Deep Pointing 4 Part 2 &17 & POINTINGFOUR-C  & 01180017001 & 2022-10-05 & 2022-10-05\\
Deep Pointing 4 Part 3 &18 & POINTINGFOUR-A  & 01180018001 & 2022-10-02 & 2022-10-03\\
\enddata
\end{deluxetable*}

Nine total filters are provided in these JADES data: F090W, F115W, F150W, F200W, F277W, F335M, F356W, F410M, and F444W.  The two medium-band filters, F335M and F410M, provide extra spectral resolution in the
3--5 $\mu$m region that samples the rest-frame optical band at high redshift.  In so doing, we reveal strong emission lines and provide measurement of the stellar continuum between them.

The total area covered by these observations is about 25~arcmin$^2$, but the area covered in the shortwave (SW) and longwave (LW) arms of NIRCam is not identical.  A sizeable portion of the area is covered by two pointings, substantially increasing the depth.  There are 2 thin uncovered strips in F200W due to the SW chip gap; these were covered in other filters and will be covered in F200W in year 2 JADES observing.  We mention that the area covered in all 9 filters is set by the intersection of the F200W and F356W footprints.

We combine the JADES data with imaging from the JEMS \citep[program 1963][]{williams2023a}.  JEMS imaged one NIRCam field in five medium-band filters, F182M, F210M, F430M, F460M, and F480M.  While not as deep as the JADES data, these filters provide additional resolution on the spectral
energy distributions and these images reveal more emission lines over a larger range of redshifts in these high-redshift galaxies.  

Table \ref{table:exposures} summarizes the exposures and achieved sensitivities. 
In all 14 filters, JADES and JEMS provide extensive pixel diversity, allowing substantial mitigation of flat-fielding errors, cosmic rays, and other pixel-level
issues.  This proved important in the JADES data, as four of the observations suffer from substantial persistence in 3 of the 8 SW detectors.  These are described more in \citet{eisenstein23r}.  Our masking mitigations are described in \S~\ref{sec:masking}.

\setlength{\tabcolsep}{1pt}
\begin{deluxetable*}{ccccccc}[htb]
\tablecaption{Image Mosaic Properties}
\tablehead{\colhead{Filter} & \colhead{Program} & \colhead{\hspace{1.5cm} 5-$\sigma$ Flux Depth} & \colhead{} &Aperture & \colhead{Exposure Time} & \colhead{Area} \\ 
\colhead{} & \colhead{ID} & \colhead{[nJy]} & \colhead{AB mag 0.3" diam.} &Correction& \colhead{90\% (20\%) [s]} & \colhead{[arcmin$^{2}$]} } 
\startdata
\multicolumn{7}{c}{JADES}\\
F090W & 1180 & 5.8 & 29.49 & 1.26 &12368 (35730) & 24.9 (6.3) \\
F115W & 1180 & 4.3 & 29.82 & 1.24 &24737 (60468) & 25.0 (6.7) \\
F150W & 1180 & 4.6 & 29.74 & 1.22 &12368 (35730) & 25.2 (7.3) \\
F200W & 1180 & 4.4 & 29.79 & 1.23 &12368 (24736) & 24.4 (8.0) \\
F277W & 1180 & 3.4 & 30.07 & 1.36 & 12368 (35730) & 25.8 (9.4) \\
F335M & 1180 & 5.9 & 29.47 & 1.43 & 12368 (24376) & 25.5 (9.8) \\
F356W & 1180 & 3.6 & 30.01 & 1.45 & 12368 (24736) & 25.6 (9.9)  \\
F410M & 1180 & 5.7 & 29.51 & 1.49 & 12368 (35730) & 25.8 (9.4)  \\
F444W & 1180 & 4.5 & 29.77 & 1.52 &  12368 (35730) & 25.8 (9.4) \\
\multicolumn{7}{c}{JEMS}\\
F182M & 1963 & 6.1 & 29.44 & 1.23 &13914 (27829) & 10.0 (7.0) \\
F210M & 1963 & 7.2 & 29.26 & 1.24 &13914 (27829) & 10.0 (7.0) \\
F430M & 1963 & 14.5 & 28.50 & 1.51 & 6957 (13914) & 9.8 (8.1)  \\
F460M & 1963 & 18.8 & 28.21 & 1.54& 6957 (13914) & 9.8 (8.1) \\
F480M & 1963 & 13.6 & 28.57 & 1.57& 13914 (27829) & 9.8 (8.1) 
\label{table:exposures}
\enddata
\end{deluxetable*}

\section{Data Processing and Mosaics}
\label{sec:mosaics}

\begin{figure*}[ht]
\centering
\includegraphics[width=7in]{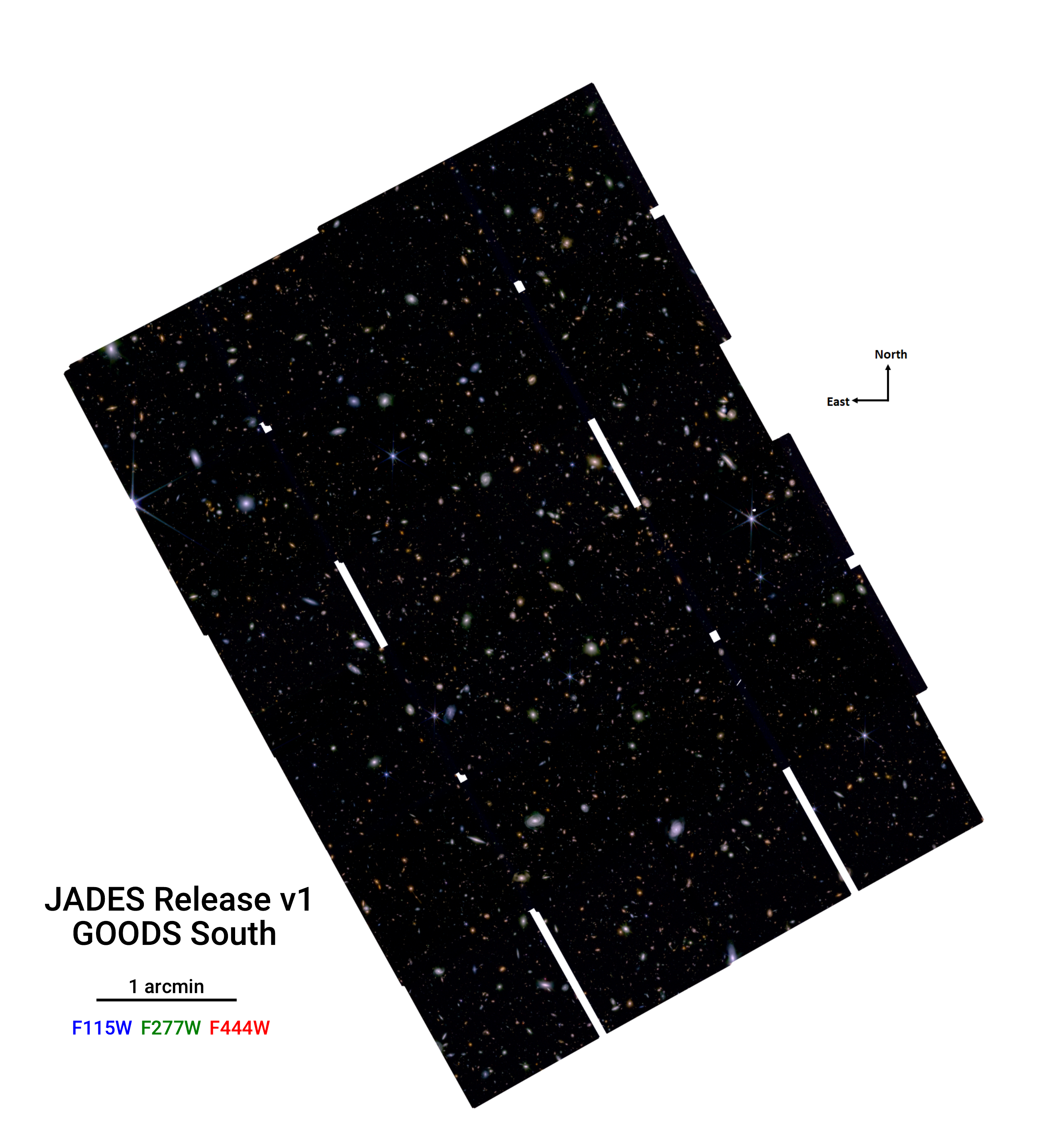}
\caption{False color image of the JADES NIRCam mosaic.
Shown are logarithmic scalings of F444W (red), 
F277W (green), and F115W (blue) filter images on a
0.3" pixel scale. The scale bar indicates one arcminute. The white regions are areas with incomplete short wavelength coverage. 
}
\label{fig:jades}
\end{figure*}

We give here an overview of how we reduce JADES and JEMS NIRCam data and construct the final mosaics that are part of this release. A more detailed discussion of the procedure including quality assessments will the presented in Tacchella et al. (in prep.). We process the images with the \textit{JWST} Calibration Pipeline (version 1.9.6) with custom steps and modifications as described below. For this first public release, we use Calibration Reference Data System pipeline mapping (CRDS pmap) 1084, which includes in-flight NIRCam dark, distortion, bad pixel mask, readnoise, superbias and flat reference files. 

\subsection{Stage 1 Detector Processing}

Stage 1 consists of detector-level corrections that are performed on a group-by-group basis, followed by ramp fitting. In this step, the data quality (DQ) arrays are initialized, saturated pixels are identified, the superbias is subtracted, the readout noise is corrected using the reference pixels, non-linearity corrections are applied, dark current is subtracted, and cosmic rays are identified. NIRCam suffers from large cosmic ray events, called ``snowballs'' \footnote{\url{https://jwst-docs.stsci.edu/data-artifacts-and-features/snowballs-and-shower-artifacts}}, which have the characteristics of a cosmic ray surrounded by a halo of pixels that have a low level of excess counts. This halo can be as large as several hundreds of pixels. The \textit{JWST} Calibration Pipeline constrains and corrects for this halo in the jump step by fitting circles that enclose the large events and expands these circles by the input \texttt{expansion\_factor} and marks them as jumps in the DQ array. We find that the identification and correction of snowballs work well in about $\sim80\%$ of the cases, with problems mostly arising from the largest hits. We run Stage 1 with the default parameters. The output of Stage 1 processing is a count-rate image in units of counts/s.

\subsection{Stage 2 Imaging Calibration}

Stage 2 processing consists of additional instrument-level and observing-mode corrections and calibrations to produce fully calibrated exposures. Specifically, this step involves flat-fielding the data and applying flux calibration that converts the images from counts/s to MJy/sr. We adopt the default values for these pipeline steps, but replace the STScI flats with super-sky flats for the JADES and JEMS LW bands. We also apply an astrometric correction which is needed for the count/s to MJy/sr conversion.

The motivation for constructing our own flats arose by finding distinct small-scale periodic patterns in the background structure of LW module mosaics of the deep JADES observations that resulted in the detection of numerous spurious sources. To mitigate this issue we created super-sky flats for the JADES and JEMS LW bands based on the JADES imaging. This procedure is detailed in Tacchella et al. (in prep) -- here we note that our tests indicate these flat fields perform at least as well as the sky-flats released to the Calibration Reference Data System (CRDS) on 4 May 2023. Recent CRDS releases provide flat fields produced in a manner similar to what was used for the JADES mosaics, both of which are an improvement over early flats which were based on ground test data and limited flight data.

Briefly, for each band the flats were generated by first constructing a mosaic and identifying sources in the image through a segmentation map.  This mosaic segmentation map was dilated and reprojected to the individual stage-1 count-rate images. Each masked rate image was divided by its median to produce an estimate of the response of the detectors to a uniform source of illumination. We then computed the median over all rate files.  For some medium bands we did not have sufficient images to produce a robust flat field; in these cases we interpolated the flat-field images from the surrounding bands.  We found that the use of these flats substantially improved the smoothness of the background in mosaiced images on small scales. 

\subsection{Custom steps post-Stage 2}

Following Stage 2, we perform several custom corrections in order to account for several features in the NIRCam images \citep{rigby22}, including the 1/f noise \citep{schlawin20}, scattered-light effects\footnote{\url{https://jwst-docs.stsci.edu/jwst-near-infrared-camera/nircam-features-and-caveats/nircam-claws-and-wisps}} (``wisps'' and ``claws'') and the large-scale background. Since all of those effects are additive, we fit and subtract them. Finally, we also updated the DQ array in order to mask additional features that led to an imprint onto the mosaics, including persistence, uncorrected wisp features and unflagged hot pixels. 

\subsubsection{1/f noise and wisp subtraction}

We assume a parametric model for the 1/f noise. We fit the source-masked, background-subtracted image (see below) with a model for the 1/f noise
\begin{eqnarray}
    D_{x,y} &= (a_{x} + b_{y,{\rm amp}} + c_{\rm amp})
\end{eqnarray}
where $a$ is a vector of coefficients of length 2048, $b_{amp}$ are four vectors  of coefficients each of length 2048 (one for each amplifier), and $c_{amp}$ are constants for each of the four amplifiers. This model is fit using GPU accelerated code.

We construct the wisp templates from the calibrated Stage-2 images. We include all SW images from PID 1180, 1181, 1210, 1286, 1837, and 1963. Specifically, similar to the flat construction, we dilate and then project the mosaic segmentation map to the individual Stage-2 images. Following this, we subtract a median background from each individual Stage-2 image and then compute a median image for each detector. We also construct a wisp mask that contains the main wisp feature. The detectors with the strongest wisp features are A3, A4, B3 and B4, while A1 shows an ``eye''-like feature on the top right. 

We then subtract a large-scale background, the scaled wisp template and the 1/f noise from the Stage-2 images. We estimate the large-scale background with the \texttt{Photutils} \texttt{Background2D} class, using the \texttt{biweight\_location} estimator to obtain the average background in sigma-clipped boxes of $256\times256$ pixels in a grid across the image. The resulting low-resolution background grid is then median-filtered over $3\times3$ adjacent boxes. After subtracting this large-scale background, we fit the wisp template in the overlapping region of the wisp mask and where no sources are present. After subtracting this normalized wisp template, we fit for and subtract the 1/f noise model as outlined above.

\subsubsection{Masking}\label{sec:masking}

We found several artefacts in the initial mosaics, which are caused by persistence, some residual wisp and claw features, and hot pixels. We visually inspected all the stage-2 images, looking for remaining artefacts and manually masked these features to clean the final images. Masks were constructed using the free and open-source raster graphics editor \texttt{Gimp}\footnote{\texttt{Gimp} version 2.10.34 available at \url{https://www.gimp.org}}. 

The persistence is thought to be originating from an observation of the Trapezium Region (Program ID 1256) taken only 30 minutes prior to the first JADES observation, which left a substantial imprint on the detectors due to its high surface brightness. Large-scale persistence is visible in the A3 and B4 detectors and some in the B3 detector, most pronounced in the F090W and F115W filters. Additionally, there was strong persistence caused by bright stars and resulting lines from the telescope slewing through possibly a star field in the preceding observations. These features were mainly present in Observations 7, 10, 15, and 18. There were some wisp- and claw-related residuals in the filters F150W and F200W, detectors A3, B1, B2, and B4, in particular in Observations 11 and 17. Furthermore, a circular feature, dubbed the ``eye'', was generally present on detector A1, in F090W and F115W. There were a few cases of uncorrected snowballs that we masked. 

All the above features (persistence, wisps, claws, and eye) were easily identifiable since they are stationary on the detector. However, there were two cases of artefacts which were moving with the background and hence could only be identified from the final mosaics. These were hexagon-shaped artefacts near bright stars, which look very similar to the telescope aperture shape. Therefore, we suggest that these are rare cases of filter ``ghosts'', which were also previously identified on \textit{HST} images\footnote{\url{https://www.stsci.edu/hst/instrumentation/wfc3/performance/anomalies}}.

Finally, in the initial mosaic, we found $\sim80$ hot pixels across all bands. Most SW hot pixels were related to large cosmic ray hits, which have not been properly masked in Stage 1. This led to bright pixels, with DQ=0. Those were not picked up by the outlier rejection in Stage 3, because the errors of those pixels were extremely large. We have addressed this by ($i$) ensuring that neighboring pixels of cosmic ray hits with fluxes in the top 99.9\% percentile are masked, and ($ii$) setting a maximum error of 0.03 in the error maps, which translate into clipping the top 99.99\% percentile. Most LW hot pixels appeared on the edges of the mosaics, consistent with having only a few images ($\approx3$) in the outlier rejection of Stage 3. We identified those hot pixels in the individual Stage-2 products and added them to those masks.

\subsection{Stage 3 -- Mosaic construction}

Stage 3 combines all the individual images and dithers into a single mosaic per filter. This step includes astrometric alignment, background matching, outlier detection, and resampling the images onto a common output grid. After a customized astrometric correction, we run Stage 3 with the default parameter values, setting the pixel scale to 0.03 arcsec/pixel\footnote{The exact pixel size of the mosaics is 0.0299947 arcsec/pixel to match the effective pixel size of the HLF HST mosaics after incorporating a slight correction to improve their registration with respect to Gaia DR2 positions.} and drizzle parameter of \texttt{pixfrac}=1 for the SW and LW images. We constructed a mosaic for each observation, i.e. a total of six sub-mosaics (Observations 7, 10, 11, 15, 17, and 18). We perform a custom background subtraction of those sub-mosaics to remove any residual background before combining them into a final mosaic, ensuring proper flux and error propagation, and perfect pixel grid alignment. In the following, we describe briefly these customized processing steps.

Our astrometric alignment process includes some modifications of the standard \texttt{jwst-pipeline} \texttt{tweakreg} procedure. The astrometric alignment of the JADES and JEMS imaging is computed relative to HST images that have been registered to Gaia DR2 \citep[G. Brammer priv. comm.,][]{gaiaDR2}.  We first construct a reference catalog of isolated, approximately round objects from these Gaia-registered HST F160W and, where F160W imaging is not available, F850LP images.  For every Stage-2 NIRCam F150W and F200W image in a given observation -- consisting of several dither positions taken after the same guide star acquisition -- we determine source positions using the SEP \citep{barbary2016a} implementation of the \texttt{SExtractor} detection code.  We then crossmatch these sources with the reference catalog and compute the rotation and offset of each level-2 image relative to the reference catalog.  Finally, we apply the median rotation and offset for this observation and guide star acquisition sequence to all level-2 images in a given observation.  For images taken in the A module with the medium band F335M and the JEMS F182M and F210M filters, we replace the default distortion maps with the nearest (in effective wavelength) wide band distortion map for that detector. After this astrometric alignment and mosaicing, we find that the relative source positions (compared to the F200W images) across the different NIRCam bands have median offsets of less than 0.07 SW pixels (2 mas) in each direction.

We estimate and subtract any remaining background in the sub-mosaics, following roughly the procedure outlined in \cite{bagley2023}. Specifically, we first generously mask sources and then measure the background in the unmasked regions using the \texttt{Photutils} \texttt{Background2D} class. We use the \texttt{biweight\_location} estimator to obtain the average background in sigma-clipped boxes of $10\times10$ pixels in a grid across the image. The resulting low-resolution background grid is then median-filtered over $5\times5$ adjacent boxes. We use the \texttt{BkgZoomInterpolator} algorithm to interpolate the filtered array and construct a smooth background model.

\section{Photometric Catalog}
\label{sec:catalog}

Using our image mosaics (\S \ref{sec:mosaics}), we generated a photometric catalog following the 
procedures described in Robertson et al. (in prep.). We provide a brief overview of the 
detection and photometry methods here, but refer the reader to Robertson et al. (in prep.) for
further details.

\subsection{Detection}
\label{subsec:detection}

To create a detection image, we produce an image of the NIRCam LW SNR as the ratio of
signal and noise images. For the signal image,
we create an inverse-variance-weighted stack of the SCI flux extensions of
F277W, F335M, F356W, F410M, and F444W images. For the noise image, we create an inverse-variance-weighted
stack of the ERR uncertainty extensions of the same images but use a median filtering to replace pixels
poorly masked by the \jwst{} pipeline. The result is a SNR image that allows for the selection
of both faint continuum sources that would be marginally detected in any one filter and strong line emitters
present in only a single NIRCam LW filter.

The subsequent detection method was inspired by the approach of \texttt{NoiseChisel} \citep{akhlaghi2015a},
which has been effective in analyzing deep images with extended low-surface-brightness features \citep[e.g.][]{borlaff2019a}.
Our pipeline implementation written in \texttt{Python} was constructed from \texttt{Astropy} \citep{astropy2022a}, \texttt{Photutils} \citep{bradley2023a}, \texttt{scikit-image} \citep{vanderwalt2014a}, and \texttt{cupy} \citep{okuta2017a} routines. In
using \texttt{Photutils}, several of the algorithms were adopted from \texttt{sextractor} \citep{bertin1996a}.
First, an initial, detection catalog of blended sources is generated by using the \texttt{Photutils}~\texttt{detect\_sources}
routine with a minimum threshold of SNR$=1.5$. A series of binary hole filling, segment expansion, erosion, and morphological
opening operations are applied to the resulting segmentation map to reduce spurious noise detections and separate objects blended
by narrow, low-surface-brightness bridges. Deblending is performed on the segmentation map using a logarithmically-scaled F200W
image by iteratively applying image denoising via median filtering, detecting peaks, and then assigning regions of the blended
segmentation to individual sources by using a watershed algorithm about the peaks. The largest segments are further deblended
using \texttt{Photutils} \texttt{deblend\_sources} using the parameters $\mathtt{nlevels}=16$ and $\mathtt{contrast}=0.1$.
We prevent objects from being deblended into sources that lie within the Gaussian-equivalent moments of its light distribution
(the ellipse defined by \texttt{Photutils} \texttt{semimajor\_sigma} and \texttt{semiminor\_sigma}). Satellites are identified by applying \texttt{detect\_sources} to the high-pass-filtered outer light distribution of extended sources. Using the segmentation
map of these identified objects, we then mask the SNR image and search for compact sources missed or removed by previous operations
by performing again \texttt{detect\_sources} on blank regions with a threshold of SNR$=3.5$. The last step in defining the
segmentation map is to apply a bright star and persistence mask determined from a simple threshold detection applied to an inverse-variance-weighted stack of all JADES images. The mask reblends shredded bright stars, includes diffraction spike segmentations from off-image stars, and covers two regions of strong persistence in the southeast corner of the mosaic.

The JADES data release includes the segmentation maps that cover the whole of the mosaic footprint with the same
pixel scale and WCS header as the images. These segmentation maps are provided in 32-bit integer format, with each
segment assigned a unique ID. The source IDs are non-contiguous to preserve consistency between object identifications
in previous internal JADES catalogs and have no special meaning.

\subsection{Photometry}
\label{subsec:photometry}

The segmentation map produced using the detection method described in \S \ref{subsec:detection} provides the sources we use in
constructing our photometric catalog.
With the segmentation map defined,
the centroids of each object are determined using the \texttt{Photutils} implementation of the \texttt{sextractor} windowed position
algorithm applied to the NIRCam LW signal image. For each object, the signal image is also used to define the Gaussian-equivalent semi-major and semi-minor elliptical sizes, their elliptical orientation on the sky, and the Kron radius. For the Kron radius, we use a
Kron parameter of $K=2.5$ and use \texttt{Photutils} to compute the Kron radius from an elliptical region with circularized
radius six times larger than the Gaussian-equivalent elliptical sizes (i.e., the \texttt{Photutils} default) while masking 
segmentation regions of neighboring sources. For small objects, and
especially small objects in regions of enhanced background flux,  this method results in artificially inflated Kron apertures that extend
well beyond their compact light distribution. We therefore limit Kron apertures to regions twice the area of the object segmentation.
As an alternative, we also compute separately apertures with a Kron parameter of $K=1.2$ and report properties for both apertures.

The JADES photometric catalogs include a variety of flux measurements in different apertures for each source, with forced photometry at the object locations determined by the detection method.
We measure circular aperture fluxes within radii of $r=[0.1", 0.15", 0.25", 0.3", 0.35", 0.5"]$ 
which correspond to CIRC1 through CIRC6 respectively. CIRC0 corresponds to the 80\% encircled energy radius determined by the PSF for each filter. We measure flux within the elliptical Kron apertures described above and the sum of the pixel
flux values within each object segmentation. This results in 10 separate flux measurements for every source in each filter.
The half-light radii for each object are measured relative to the total Kron flux in each band.

Two kinds of flux uncertainties are reported for each of the JADES NIRCam aperture
flux measurements. First, we measure directly the
uncertainties from the \jwst{} pipeline ERR extension of our mosaics for each aperture. Second, we compute random aperture-based uncertainties that account for the correlated-noise contribution to the flux error 
\citep[e.g.,][]{labbe2005a,quadri2007a,whitaker2011a,skelton2014a,whitaker2019a}. 
For a range of aperture sizes, we measure the spread
values in blank areas of each mosaic at 100,000 random locations. Since the image depth can vary dramatically over the field,
especially for the HST mosaics, these random apertures are collected into percentiles based on exposure time at each location, and then
the
power-law scaling between the RMS of counts and aperture size is fit for each percentile of apertures and recorded. These scalings
are used to determine the sky noise contribution to the flux uncertainty of each object for each aperture size,
with the sky and source Poisson uncertainties
added in quadrature to provide the total uncertainty in counts. Combining with the source flux in counts and converting to flux units, the 
source flux and total uncertainty for every object are then recorded for every aperture reported in the catalog. For the Kron fluxes,
the circularized radius of the Kron aperture is used to determine the sky noise contribution from the measured random-aperture uncertainty
scaling relations.

\begin{figure}[ht]
\centering
\includegraphics[width=8cm]{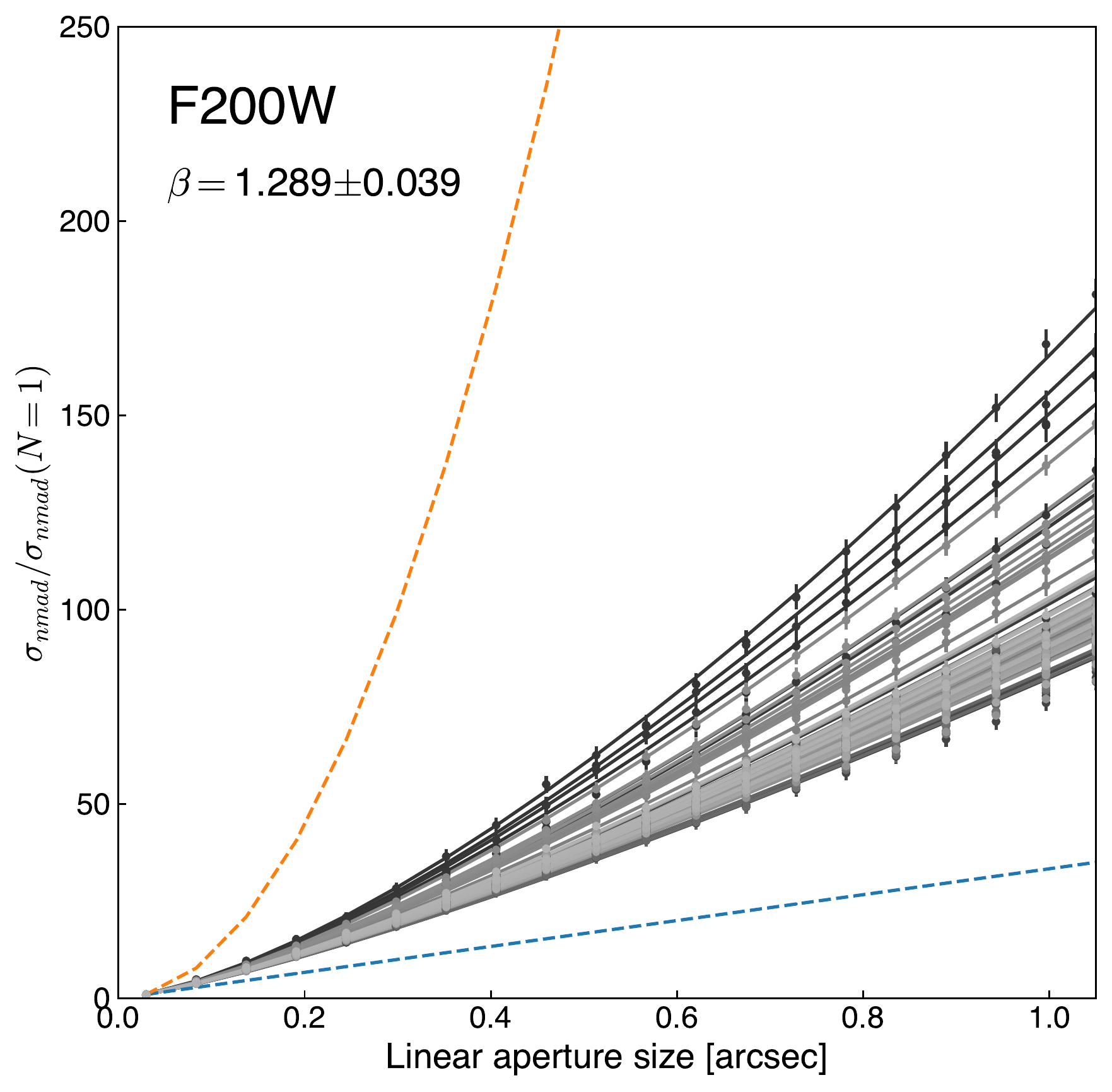}
\caption{Scaling of blank-sky RMS flux within random apertures
as a function of aperture diameter. Shown are 100 scalings of the
image noise, normalized to the single pixel noise, with 
linear aperture size for the F200W mosaic where the random apertures are split
into percentiles in exposure time. The typical power law
scaling is $\beta=1.29 \pm0.04$, compared with the idealized
$\beta=1$ for uncorrelated pixels (blue line) and worst-case $\beta=2$
for perfectly correlated pixels (orange line).}
\label{fig:nmad}
\end{figure}

Figure \ref{fig:nmad} shows an example of the scaling between the RMS noise measured in 
random apertures as a function of aperture size for the F200W image.
A scaling is measured for each subset of
1000 random apertures in each of the 100 groups split by exposure time percentiles.
The index of the power-law scaling is typically $\beta\approx1.3$ for our images,
which is a measure of the degree of pixel covariance in our mosaics. For reference,
for perfectly uncorrelated pixels $\beta=1$ and for perfectly correlated pixels $\beta=2$.
Pixel correlations are difficult to avoid on projected and recombined images, but as
judged by these scalings the overall level of correlated noise present in our images
is comparable to or better than other space-based datasets \citep[e.g.,][]{skelton2014a,whitaker2019a}.

For aperture corrections of NIRCam filters, we use the model PSFs (mPSFs) from \cite{Ji_etal2023} constructed by mosaicing
WebbPSF models repeatedly over the field identically to our exposure mosaics and then measuring the average PSF. Circular aperture
corrections are determined by computing the encircled flux with radius, while Kron aperture corrections are computed by summing
the mPSF within elliptical apertures for each source. For the HST aperture corrections, we build empirical PSFs using the 
\texttt{Photutils} \texttt{EPSFBuilder} algorithm using a list of stars compiled from the HLF \citep{whitaker2019a} and those
determined by the star locii in color-magnitude-size relations.

\subsection{Common PSF-Matched Images and Photometry}
\label{sec:common_psf}

For color selection and spectral energy distribution fitting, measuring source photometry
on common PSF-matched images can ameliorate the effect of varying resolution with wavelength on
color. 
For our PSF-matched images, we use the NIRCam F444W mPSF as our target convolutional
kernel. We generate PSF-matched images for all HST and JWST filters except the
target image F444W and the NIRCam LW F430M, F460M, and F480M images.
To generate common PSF images, we use a Wiener filtering method to regularize the
convolution by the pixel-level Poisson power $\sigma^2$ in blank regions
and apply a low-pass filter
to reduce high-frequency noise from the HST ePSF and JWST mPSF models.
Mathematically, our convolutional kernel is then
\begin{equation}
K(f) = \tilde{P}_t(f) \frac{ \tilde{P}_i^{\star} \left[S(f)-\sigma^2\right]}{|\tilde{P}_i|^2 (S(f) - \sigma^2) + \sigma^2}
\end{equation}
\noindent
where $\tilde{P}_t$ is the Fourier transform of the target (F444W) PSF, 
$\tilde{P}_i$ is the Fourier transform of the PSF for the image we are convolving,
$\sigma^2$ is the Poisson per-pixel noise power measured in blank regions, 
and $S(f)$ is the power spectrum of the total (signal plus noise) image.
The $\star$ indicates complex conjugation and $||^2$ represents the absolute
square magnitude. This Wiener filtering kernel reduces to a simple deconvolution with
$P_i$ and convolution with $P_f$ for a theoretical noise-free image.
We apply a Cosine Bell (\texttt{Phoutils} \texttt{CosineBellWindow}) to the
kernel in Fourier space before multiplying $K(f)$ by the Fourier transform
of the signal image. The parameter $\alpha$ of the Cosine Bell represents
the fraction of array values tapered by the window, and when applied to our
kernel larger values of $\alpha$ pass higher frequencies. We set $\alpha=0.9$
for JWST NIRCam SW filters, $\alpha=0.45$ for JWST NIRCam LW filters,
$\alpha=0.3$ for HST ACS filters, and $\alpha=0.4$ for HST WFC3 filters.
The common PSF-matched images are inspected for numerical artifacts and ePSFs
constructed from stars in the images to test the quality of the PSF matching.
We find that the ePSF of each common PSF-matched image has a RMS width within
10\% of the F444W mPSF and the encircled energy curves agree to within $\sim1\%$
by $r=0.3"$.

\section{Data Properties}
\label{sec:data_properties}

With our JADES pipeline image and catalog data products provided
as part of this initial release, we present below several characteristics of the
data properties and quality. While many  detailed tests of the images, photometry,
and catalogs are conducted by Tacchella et al. (in prep.) and Robertson et al. (in prep.),
we provide some examples of the data here as a demonstration of its quality and content.
We examine the image exposure time distributions
(\S \ref{subsec:texp}), noise properties
(\S \ref{subsec:noise}), 
and source count distribution with
magnitude (\S \ref{subsec:counts}), 
then present a comparison of JADES to previous deep infrared observations of the field, including a detailed comparison to CANDELS photometry
(\S \ref{subsec:ir_comparison}).

\subsection{Exposure Times}
\label{subsec:texp}

\begin{figure}[ht]
\centering
\includegraphics[width=8.5cm]{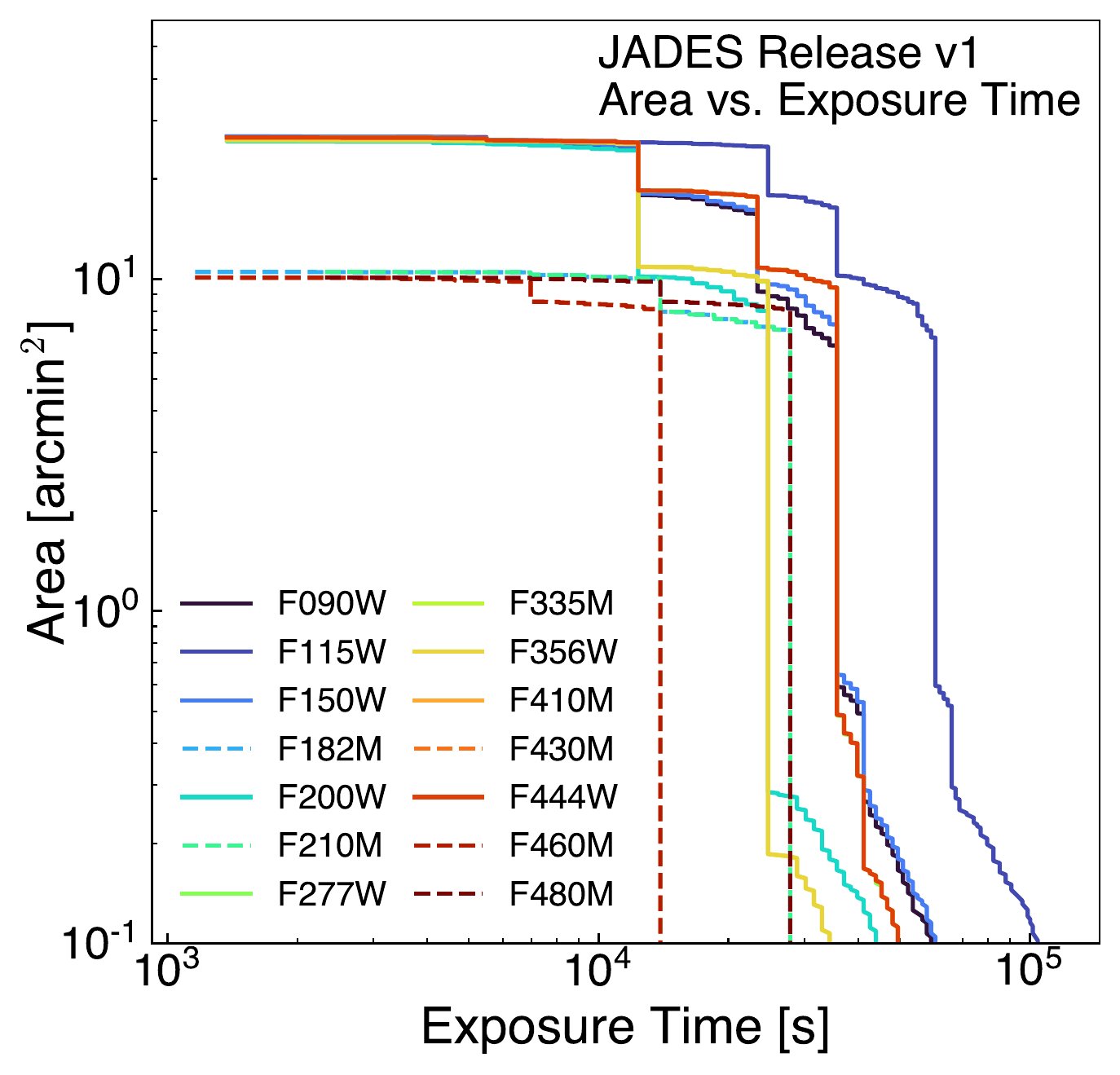}
\caption{Distribution of exposure times in the JADES NIRCam F090W, F115W, F150W, F200W, F277W, F335M, F356W, F410M, F444W and JEMS NIRCam F182M, F210M, F430M, F460M, and F480M images. Exposure times range from
$t\approx13900$s over $Area\sim10$ arcmin$^2$ in F460M to $t\approx13700-60500$s over
$Area\sim5-25$ arcmin$^2$ in F115W.}
\label{fig:texp}
\end{figure}

The exposure times of the JADES and JEMS imaging vary among the NIRCam bands.
Figure \ref{fig:texp} shows the area covered by the JADES F090W,
F115W, F150W, F200W, F277W, F335M, F356W, F410M, and F444W 
and JEMS F182M, F210M, F430M, F460M, and F480M images against the
exposure time.
The JEMS images cover approximately $10.1$ arcmin$^{2}$ at
depths of $t\ge13914$s in the F430M and F460M bands and $t\ge27829$s
in F182M, F210M, and F480M bands.
The JADES images all cover at least $25$ arcmin$^{2}$ to $t\ge12300$s.
F356W additionally covers at least $10$ arcmin$^{2}$  to $t\ge24736$s.
F115W reaches $t\ge35731$s over at least $15$ arcmin$^{2}$ and 
$t\ge60468$s over at least $6.7$ arcmin$^{2}$.
The other JADES bands reach 
$t\ge23300$s over at least $17$ arcmin$^{2}$
and  $t\ge35000$s over at least $6.3$ arcmin$^{2}$.

\begin{figure*}[ht]
\centering
\includegraphics[width=7in]{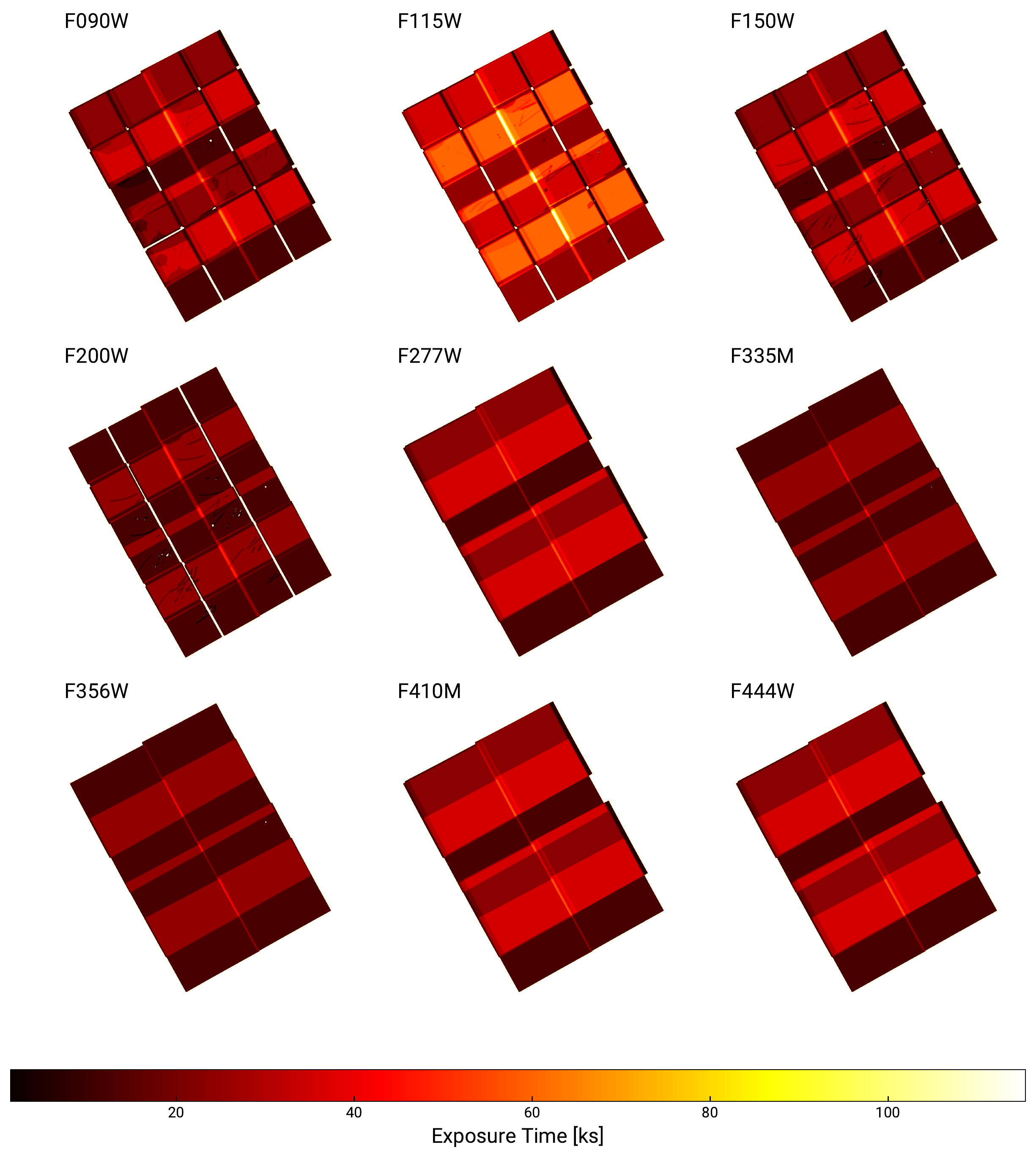}
\caption{Exposure maps for the JADES NIRCam mosaics.
Shown are the JADES F090W, F115W, F150W,
F200W, F277W, F335M, F356W, F410m, and F444W 
filter exposures.
Bright regions indicate areas of increased depth,
as reflected by the color bar scale.}
\label{fig:exp_maps}
\end{figure*}

Figure \ref{fig:exp_maps} shows the 
JADES exposure maps in the
F090W, F115W, F150W, F200W, F277W,
F335M, F356W, F410M, and F444W filters.
The 
exposure maps reveal the
mosaic tiling of the JADES observations
in the NIRCam SW and LW filter.
For the SW filters, the masking of persistence
and other data quality issues is apparent
in each of the F090W, F115W, F150W, and F200W
maps. The map morphology also shows
the extra exposure time invested in F115W
as bright regions.

\subsection{Noise Properties}
\label{subsec:noise}

The photometry measured at random locations used in \ref{subsec:photometry} to define
the aperture flux uncertainties in the catalogs reveal the noise
properties of the images. While a detailed image quality analysis will be provided
by Tacchella et al., in prep., we provide a few example measurements of
image noise properties that demonstrate the high quality of our reduction method.

\begin{figure}[ht]
\centering
\includegraphics[width=7.5cm]{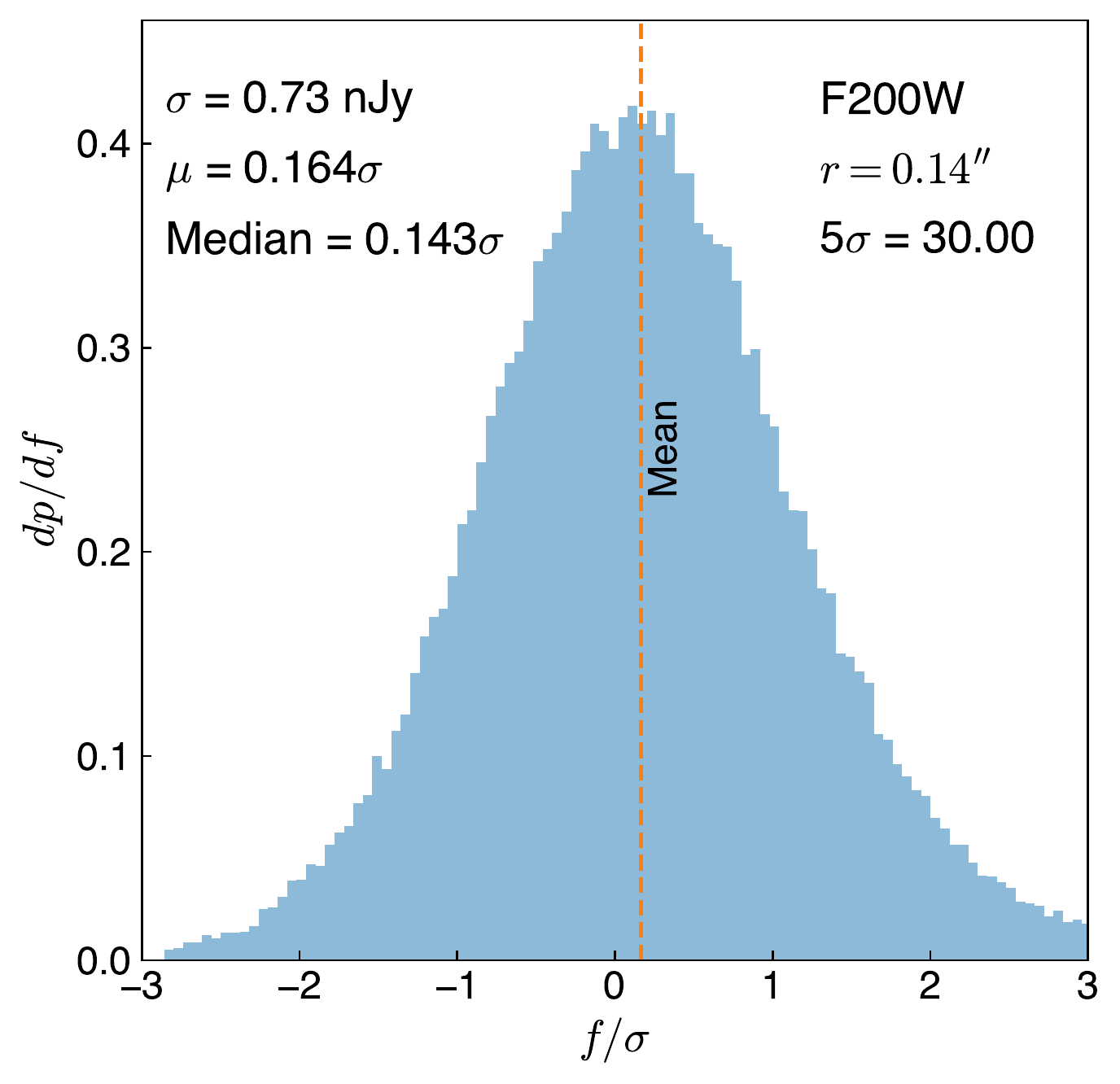}
\caption{Example blank-sky fluxes measured in randomly-distributed 0.14" radius apertures. Shown is the distribution
of random aperture fluxes measured on the F200W mosaic after applying $5\sigma$-clipping. The raw statistics of the 
distribution are listed in the figure. The median background
flux is $<0.2$nJy. After aperture correction, the RMS flux uncertainty in $r=0.14$" apertures is $\sim4.3$nJy.}
\label{fig:aperture_hist}
\end{figure}

\begin{figure}[ht]
\centering
\includegraphics[width=8.2cm]{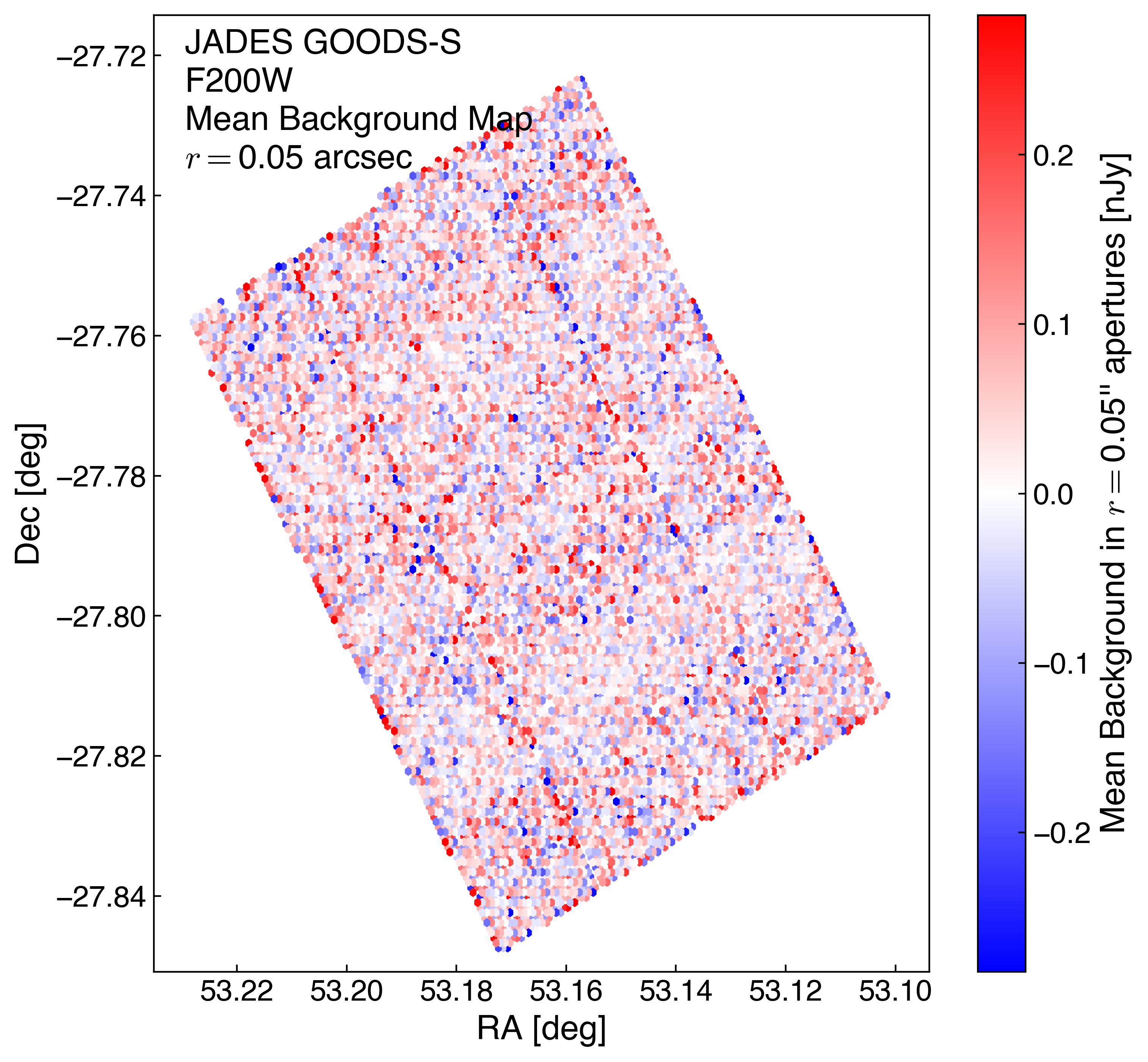}
\caption{Spatial distribution of the local mean background
in the JADES F200W image, measured in $r=0.05$" radius apertures. The mean background over the whole image is $\lesssim0.1$ nJy, with local variations in the mean background everywhere
$\lesssim0.3$ nJy.}
\label{fig:bkg_map}
\end{figure}

\begin{figure}[ht]
\centering
\includegraphics[width=8.2cm]{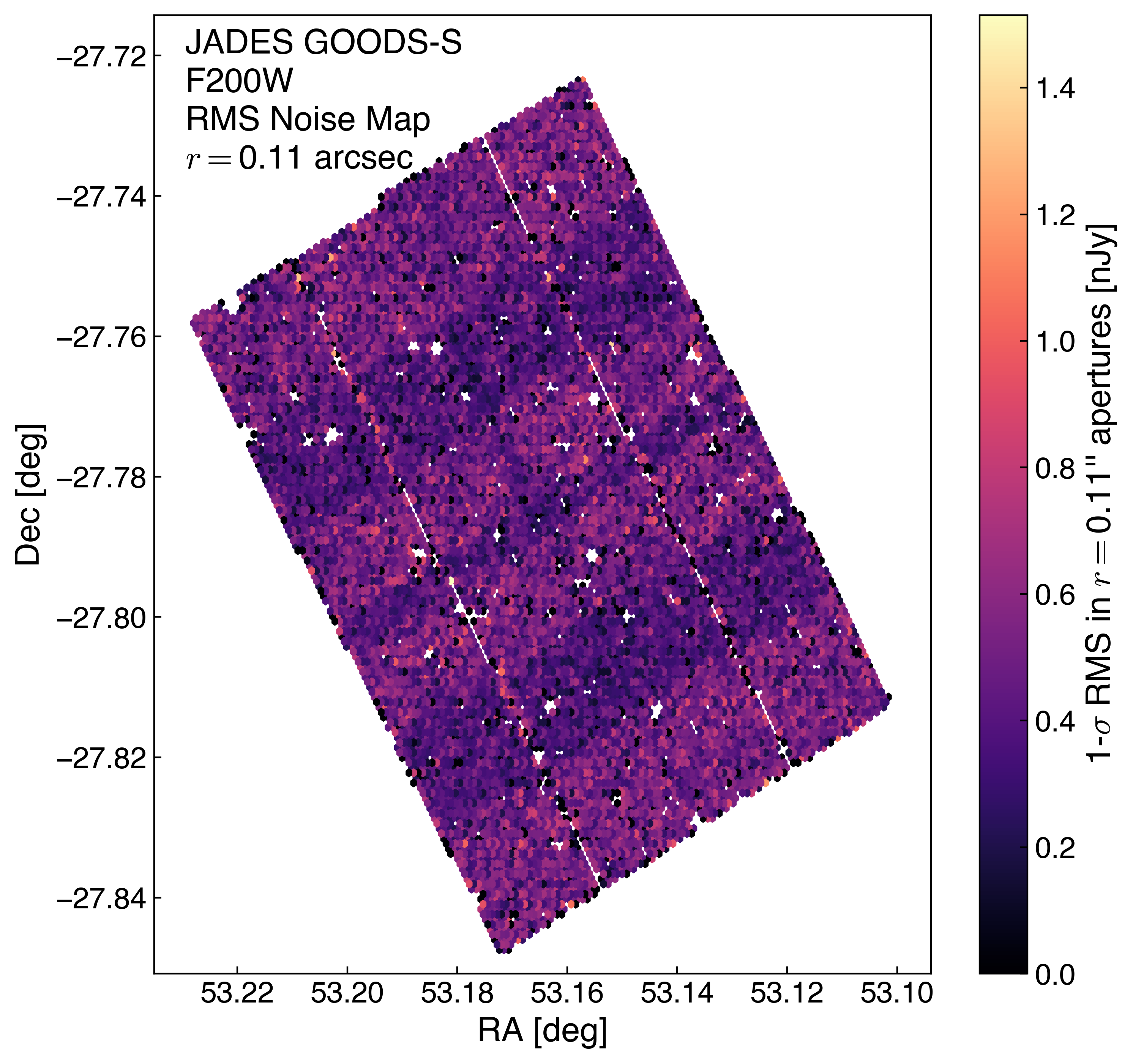}
\caption{Map of the RMS noise across the F200W image, measured
in randomly-distributed $r=0.11$" radii apertures. The
spatial variations in the random aperture RMS flux reflect the
exposure time map of F200W.
}
\label{fig:noise_map}
\end{figure}

Figure \ref{fig:aperture_hist} shows the histogram of flux values for $r=0.14"$ radius 
circular apertures measured in blank regions of the F200W image, before aperture 
correction. The median background flux measured after $5\sigma$-clipping
is $f\approx0.1$ nJy, demonstrating
control of the global background subtraction to $\sim$33.8 AB. The RMS flux
is $\sigma = 0.73$ nJy, corresponding to a $5\sigma$ depth of $30$AB (before aperture
correction).
These random apertures can be binned spatially to measure the location-dependent
mean and RMS noise across each image. Figure \ref{fig:bkg_map} shows variations
in the mean background measured in small $r=0.05"$ circular apertures across
the F200W image. Regions with lower exposure time have larger spatial variations,
especially near the image edges and intermodule regions of the mosaic, but no
strong gradients are present within regions of comparable exposure depth and
the background maintains its zero mean spatially well. The RMS of the aperture
flux values can also be visualized, as shown in Figure \ref{fig:noise_map}
that shows the RMS sky noise measured in $r=0.11"$ circular apertures across
the F200W image. The regions of enhanced exposure are apparent as regions with
lower RMS noise, and the increased noise in intermodule regions with fewer
contributing exposures is also visible. Overall, the noise map shows no
gradients in regions of comparable exposure time and we conclude that the
JADES images have quite uniform noise properties. 
We note that Figures \ref{fig:aperture_hist}, \ref{fig:bkg_map}, and
\ref{fig:noise_map} all intentionally use differing aperture sizes to
illustrate the richness of the statistics available for the images. 
In practice, as part of our data validation, such plots are automatically
created and compiled for each aperture size and each filter. These
hundreds of plots are visually inspected to help validate our combined
pipeline, identify potential issues in our reductions, and cross check
our data quality between subsequent versions of image 
reductions and catalogs.

\subsection{Source Counts}
\label{subsec:counts}

\begin{figure}[ht!]
\centering
\includegraphics[width=7.5cm]{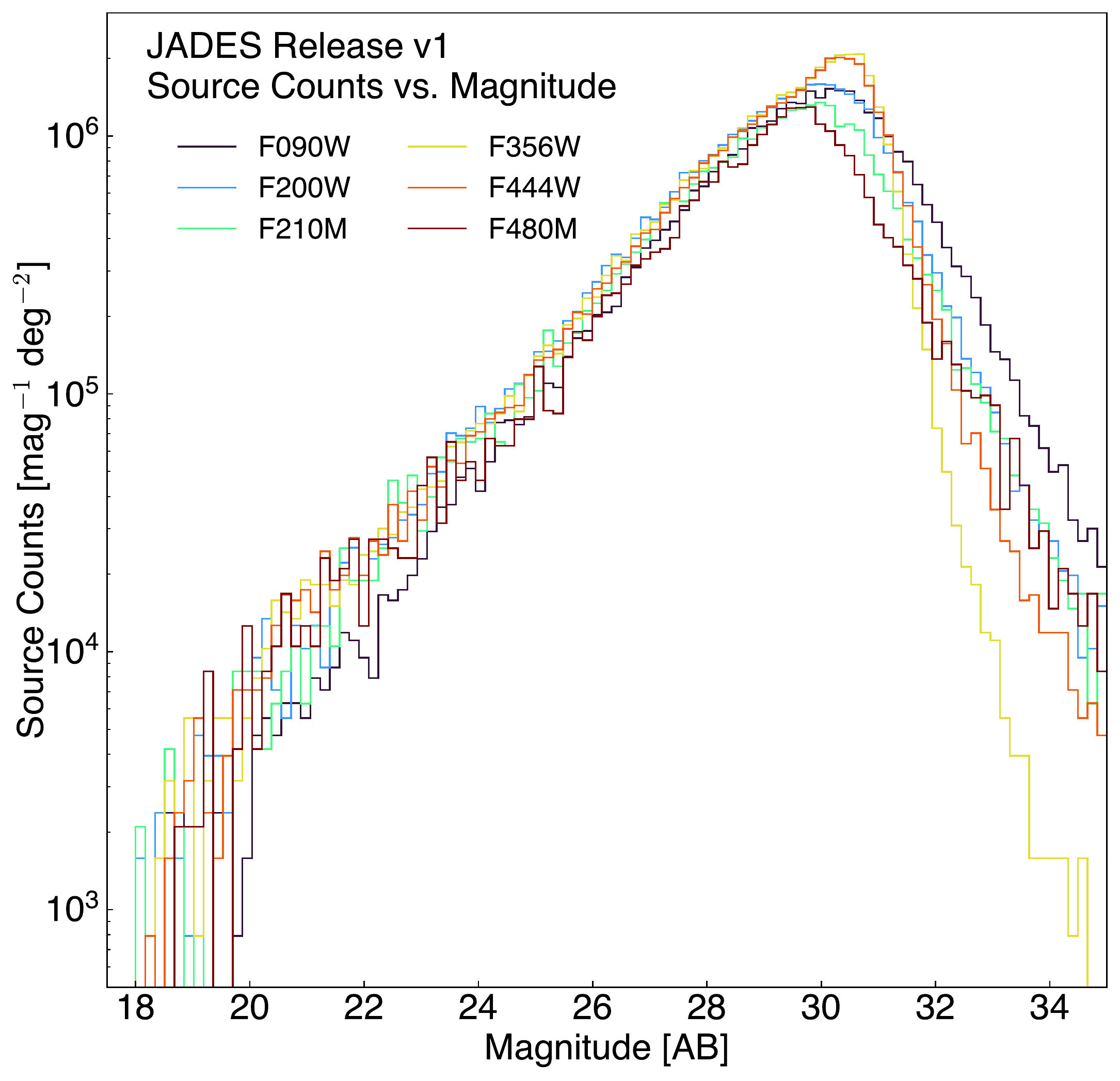}
\caption{Source counts of objects in the JADES v1 catalog release as a function of object AB magnitude. Shown are the source count distributions for the JADES F090W, F115W, F356W, and F444W bands and the JEMS F210M and F480M bands. The sources plotted are pulled from the
JADES multiband detection catalog with forced photometry in each band, such that these
histograms contain common sources where their corresponding images overlap spatially.
These bands were chosen to cover the range of exposure time vs. area distributions (see Figure \ref{fig:texp}).
}
\label{fig:source_counts}
\end{figure}

From the JADES NIRCam catalogs, we can construct the distribution of
sources as a function of their flux in different bands. The source 
count distribution will depend on the image filter owing the
color of sources, which induces a horizontal shift in the
distribution, and the image depth that controls the fall
off in source densities toward faint flux levels.
Figure \ref{fig:source_counts} shows the number of objects
per unit magnitude per square degree as a function of
AB magnitude for Kron apertures measured in the F090W, F200W, F210M (JEMS),
F356W, F444W, and F480M (JEMS) images.
The same sources
are used in each histogram, with forced photometry
applied to the images at source locations as described
in \S \ref{subsec:photometry}. The relative
depth of the filter images (see Table \ref{table:exposures})
are reflected in the location of the histogram peak, and the
distribution of exposure times (see Figure \ref{fig:texp})
affect how rapid the source
counts decline once completeness becomes poor. 


\subsection{Comparison to other GOODS-S Infrared Data}
\label{subsec:ir_comparison}

\begin{figure}[ht!]
\centering
\includegraphics[width=8.5cm]{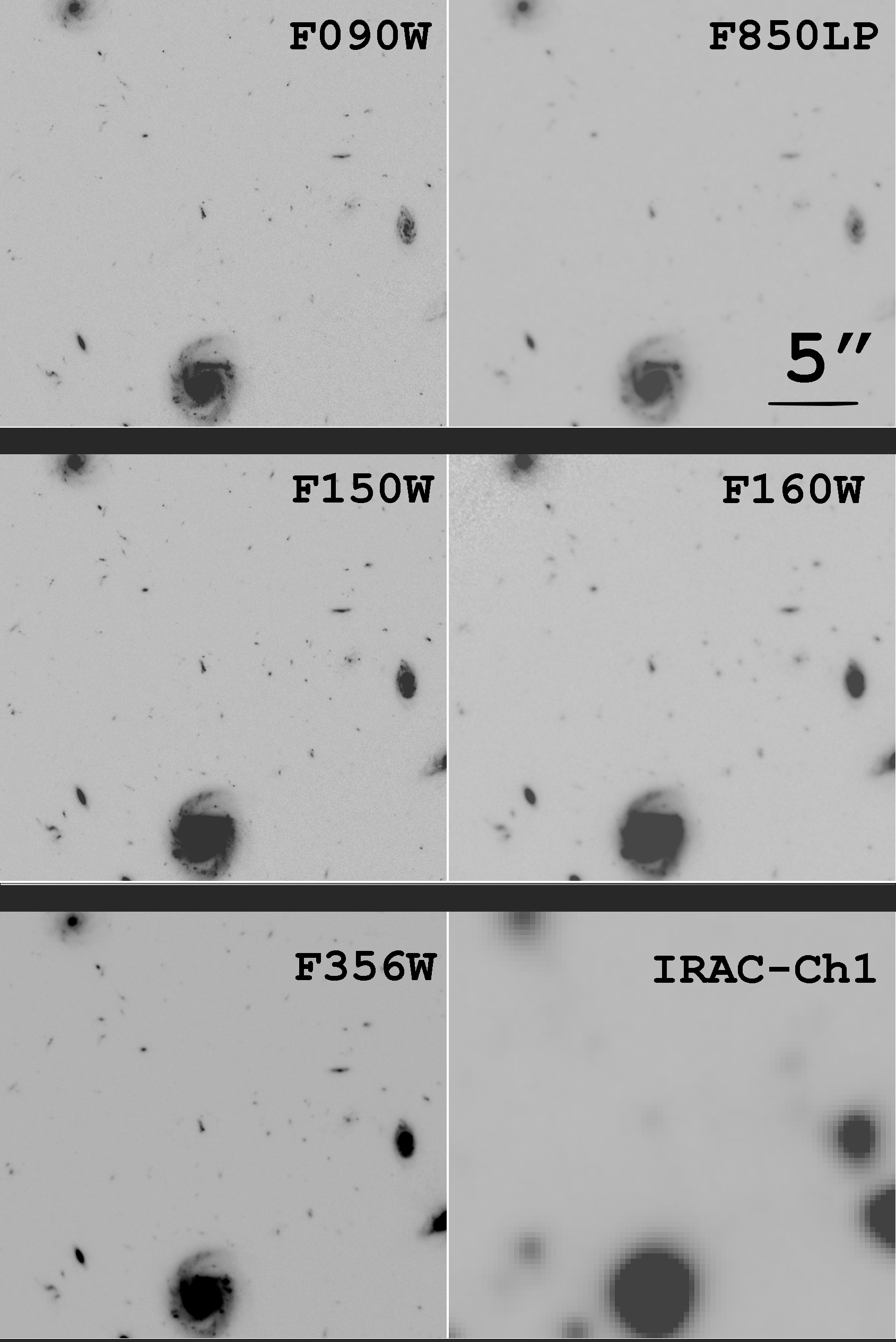}
\caption{A comparison of JWST imaging to that of HST and Spitzer, using the same small region of the HUDF.
({\it top}) The NIRCam F090W image on the left, compared to the HST F850LP image \citep[116 hrs][]{whitaker2019a} on the right.
({\it middle}) The NIRCam F150W image on the left, compared to the HST F160W image \citep[65 hrs][]{whitaker2019a} on the right.
The JWST imaging is notably sharper than HST, and mildly deeper than even these very long HST exposures. 
({\it bottom}) The NIRCam F356W image on the left, compared to the
Spitzer IRAC Channel 1 3.6~$\mu$m image \citep{Stefanon_etal2021} on the right.
JWST is far deeper and sharper than the deepest Spitzer data.
}
\label{fig:hst_irac_comp}
\end{figure}

The data in this release cover portions of the heavily observed
GOODS-S field, which provides the HST data that comprises the HLF images we
utilize and opportunities for comparisons between published catalogs and
our JADES release catalog. 
A visual impression of the comparison of JADES imaging to that of the deep portion of the HUDF is shown in Figure \ref{fig:hst_irac_comp}.  Here we compare F090W to the HST ACS F850LP imaging and then F150W to the HST WFC3 F160W imaging.  
Unsurprisingly, the JWST images show a substantially sharper point spread function, and JADES is mildly deeper.  Of course, JADES is providing a much wider field, as this deep ACS and WFC3 HST data are only about 11 and 5 arcmin$^2$, respectively.

Beyond 1.6~$\mu$m, JWST is unparalleled, as shown by the comparison of F356W to the imaging at 3.6~$\mu$m with Spitzer IRAC \citep{Stefanon_etal2021}.  The HUDF is one of the very longest exposures ever with Spitzer, but clearly the sharpness and depth of the JWST imaging are far superior.

\begin{figure*}[p]
\centering
\includegraphics[width=6in]{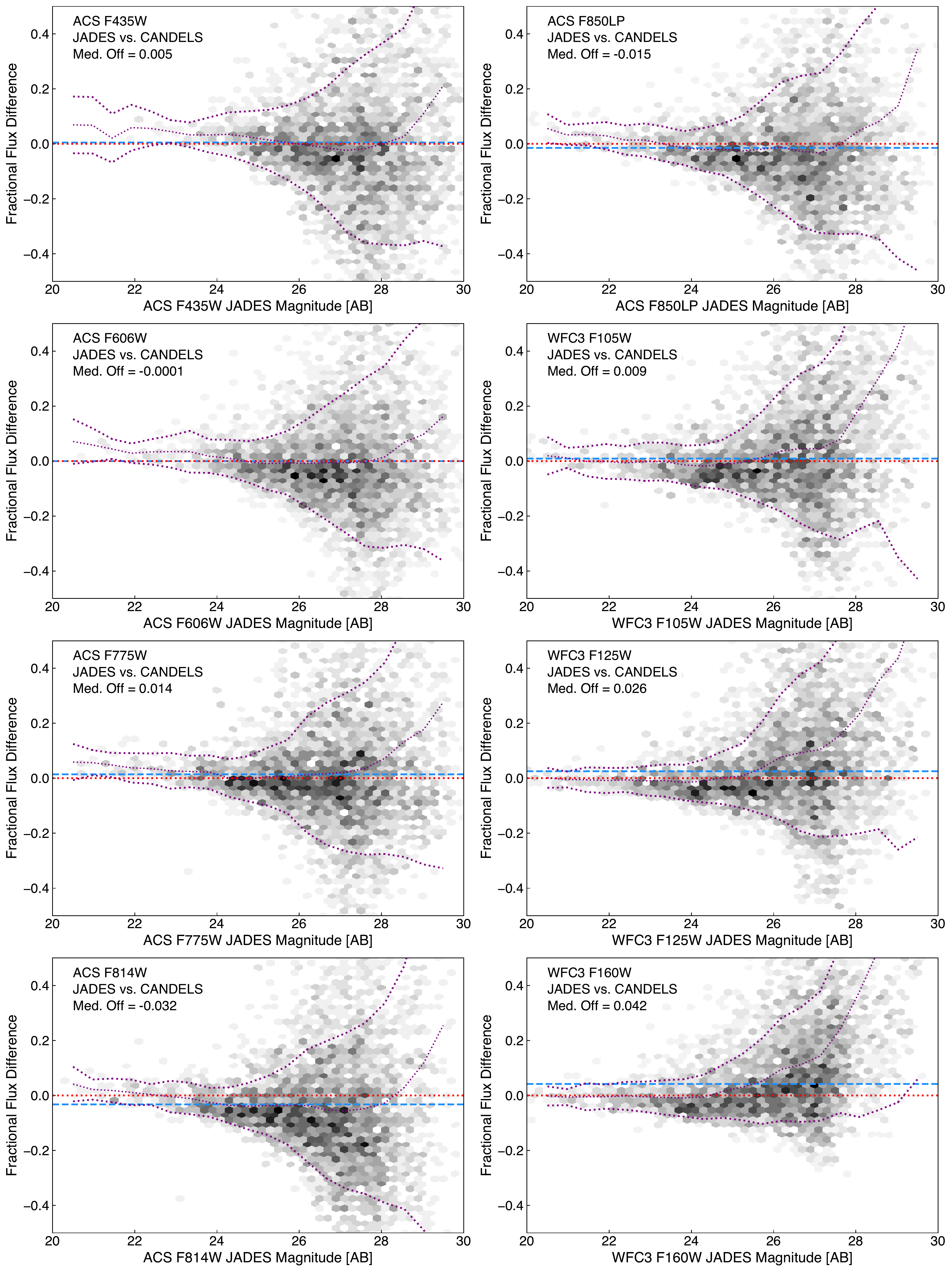}
\caption{Comparison between JADES photometry on the HLF HST images and the photometry reported by CANDELS \citep{guo2013a}. Shown are histograms of the fractional difference between the JADES Kron-aperture photometry and the total flux reported in v1.0 of the CANDELS multiwavelength catalog, relative to the JADES flux as a function of the JADES AB magnitude. The histograms are measured for objects in the CANDELS catalog with a counterpart in the JADES catalog matched within 0.5". Shown are histograms for ACS F435W, F606W, F775W, F814W, and F850LP and WFC3 F105W, F125W, F160W. We also report the fractional flux median offset between CANDELS and JADES for objects brighter than the $5\sigma$ extended object depth reported by \citet{grogin2011a} for each CANDELS band (blue dashed line). We show zero fractional difference with a dotted red line. The dotted purple lines show the running median and $\pm1\sigma$ spread of each distribution.}
\label{fig:jades_vs_candels}
\end{figure*}

To create a quantitative check on our photometry, we focus on the imaging from the 
CANDELS survey \citep{grogin2011a,koekemoer2011a}, which imaged a broader area in GOODS-S.
We use HST F435W, F606W, F775W, F814W, and F850LP ACS images and
F105W, F125W, and F160W WFC3 images that derive from the CANDELS
program for measuring blue photometry
for the JADES sources. 
We match sources between the CANDELS catalogs created by \citet{guo2013a}
and the JADES catalogs, and then compare the measured photometry in 
each HST filter. Figure \ref{fig:jades_vs_candels} shows the histogram
of fractional flux difference between the Kron photometry of
\citet{guo2013a} and JADES as a function of source AB magnitude in
the eight HST filters available in the CANDELS v1 catalog. The
relative agreement is excellent, and we compute a median offset
in fractional flux difference between CANDELS and JADES of $<0.0001$
for F606W, degrading to $0.042$ in WFC3 F160W. The median offsets
are computed for objects brighter than the $5\sigma$ flux limit
of CANDELS GOODS-S Wide reported by \citet{grogin2011a}. We note that
we do not use WFC3 in our SED fitting, and usually only combine JWST NIRCam
with HST ACS data in our analyses.

\begin{figure}[ht!]
\centering
\includegraphics[width=8.5cm]{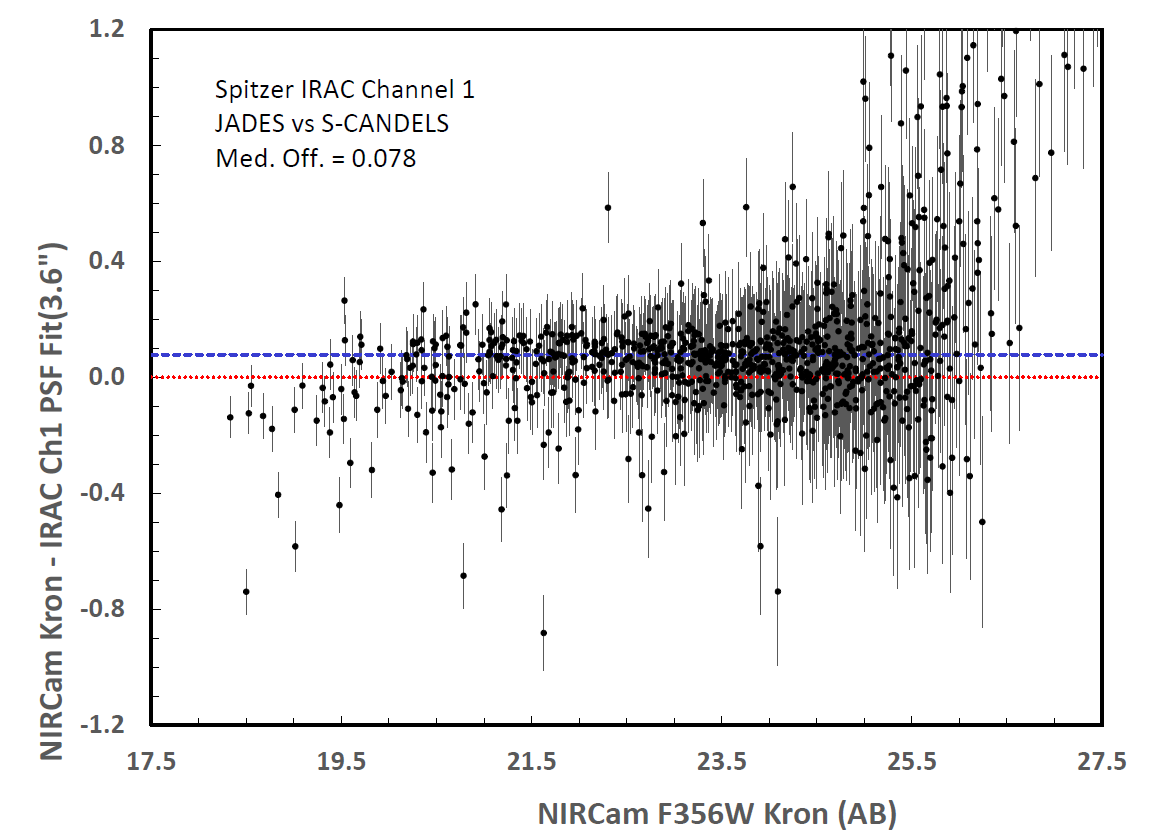}
\caption{Comparison of JADES F356W photometry with IRAC Channel 1 photometry reported in \cite{ashby2015}. The
JADES AB magnitude for Kron radii is plotted against the difference between the JADES AB magnitudes and IRAC data using a PSF fit computed over a 3.6" diameter area. Sources were required to match with 0.2" (see text). The blue dashed line indicates the median offset between JADES and IRAC data. Zero difference between the two data sets is
indicated by a dotted red line.}
\label{fig:jades_vs_irac}
\end{figure}

As a further check on our photometry, we compare NIRCam F356W magnitudes in Kron apertures to IRAC Channel 1
magnitudes using PSF fits to data in 3.6" apertures presented in \cite{ashby2015}. IRAC Channel 1 has a very similar bandpass to NIRCam's F356W filter but has very different spatial resolution with $\sim 400$ NIRCam long wavelength pixels being equivalent
to one IRAC pixel. Figure \ref{fig:jades_vs_irac} shows this comparison with a relatively small offset between the two datasets. Sources
from JADES were matched to IRAC sources requiring object centroids to agree within 0.2". The positional requirement is tighter
than for the comparison to ACS because confusion in the IRAC data can cause light barycenters to be dragged off the main object.
The large scatter is partly due to the signal-to-noise ratios in the IRAC data and partly because of the resolution difference. 

\section{Photometric redshifts}\label{sec:photoz}

We measure photometric redshifts for the resulting source catalog using the template-fitting code {\tt EAZY} \citep{brammer2008}. {\tt EAZY} combines a set of user-defined  galaxy templates to fit the observed photometry for each source across a redshift grid. We adopt the value corresponding to the overall $\chi^2(z)$ minimum as our photometric redshift ($z_{phot}$) and use the output $P(z)$ surface to estimate uncertainties ($P(z) = \exp{[-\chi^2(z) / 2]}$). For this catalog, we follow the fitting procedure, including the templates and parameters used as described in \cite{hainline23}. 

Because of the large range of redshifts and galaxy sizes in the catalog, we perform fits using the Kron fluxes and uncertainties estimated from the PSF-matched mosaics. We fit to the JADES JWST/NIRCam F090W, F115W, F150W, F200W, F277W, F335M, F356W, F410M, and F444W photometry. At shorter wavelengths, we combined these data with the HST/ACS F435W, F606W, F775W, F814W, and F850LP fluxes. For a portion of the field where there was overlap, we also added the JEMS JWST/NIRCam F182M, F210M, F430M, F460M, and F480M medium-band fluxes. We opted to not use the HST/WFC3 photometry for the fitting because of the poorer spatial resolution and lower SNR.

\begin{figure}[ht!]
\centering
\includegraphics[width=8.5cm]{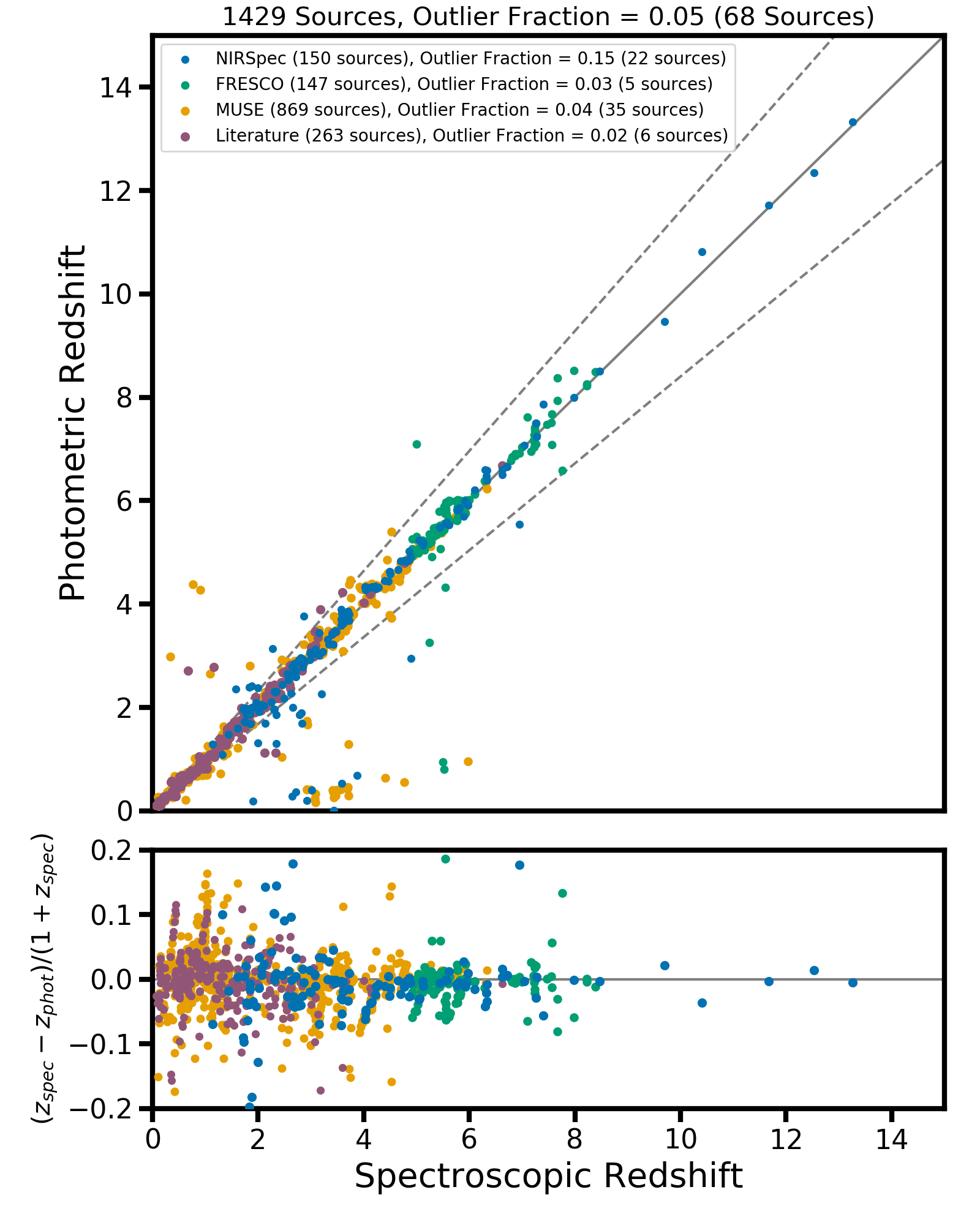}
\caption{Photometric redshifts plotted against spectroscopic redshifts. The photometric redshifts were calculated using the PSF-convolved mosaics and Kron elliptical aperture fluxes. We color the points by whether they're in the broad literature of GOODS-S spectroscopic redshifts (dark pink), whether they're derived from MUSE observations (gold), FRESCO observations (green), or NIRSpec observations (blue). We find an overall outlier fraction of 5\%, $\langle z_{spec} - z_{phot} \rangle = 0.05$,  $\sigma_{NMAD} = 0.024$. The cloud of points at $z_{spec} = 3 - 4$ and at $z_{phot} = 0 - 1$ are those sources where the Lyman break has been mistaken as the Balmer break in the fit.}
\label{fig:photoz_vs_specz}
\end{figure}

\begin{figure}[ht]
\centering
\includegraphics[width=8.5cm]{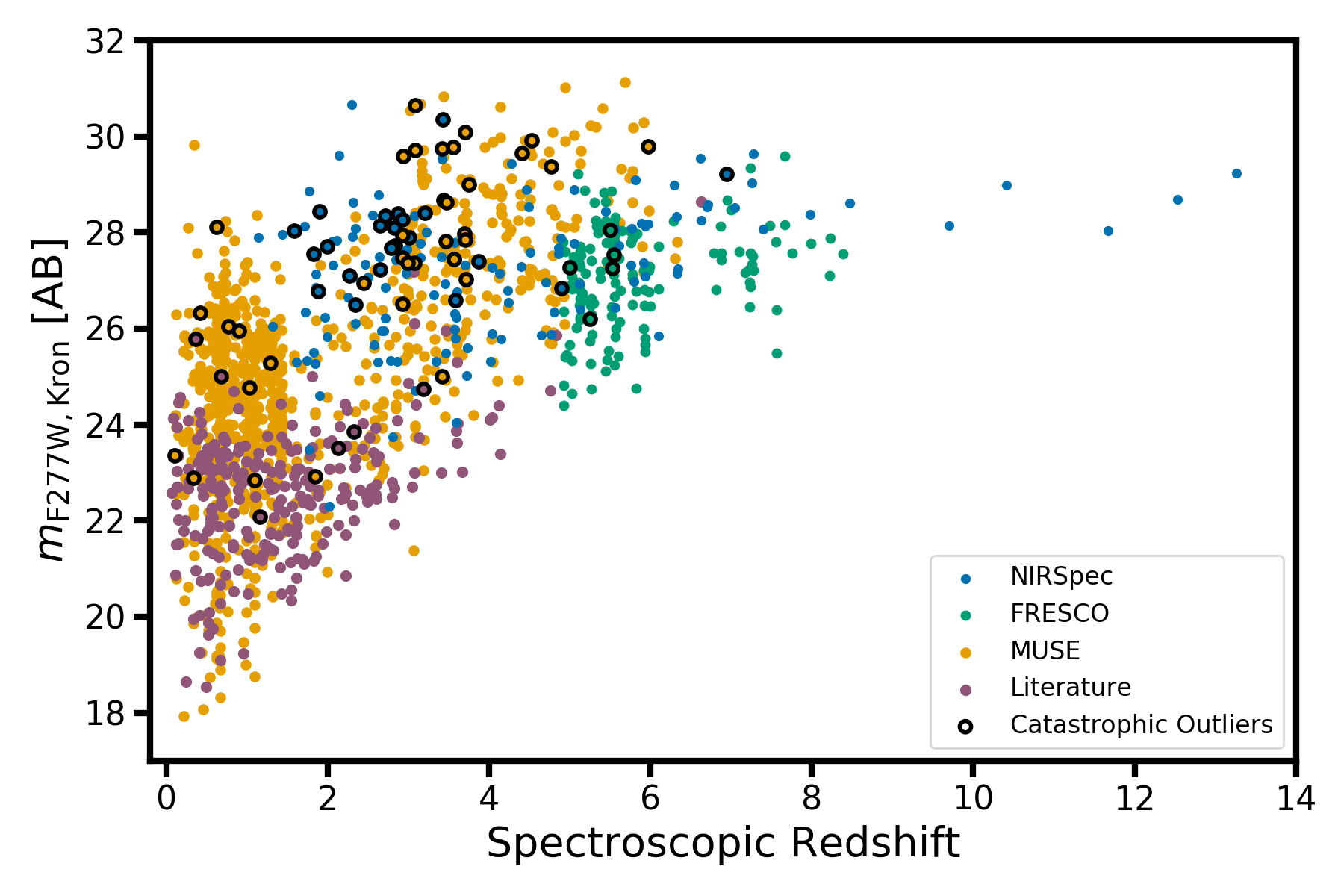}
\caption{F277W Kron magnitude plotted against spectroscopic redshifts. We plot the objects in each category of spectroscopic redshifts with the same colors as in Figure \ref{fig:photoz_vs_specz}. We also plot catastrophic outliers in the photometric redshift vs. spectroscopic redshift plot with black outlines on the points. We can see the observed F277W magnitudes for the spectroscopic redshift comparison sample extends across a wide range, and that many of the $z > 3$ catastrophic outliers are faint sources.}
\label{fig:m277w_vs_specz}
\end{figure}

To demonstrate the accuracy of the estimated photometric redshifts we collected spectroscopic redshifts from the literature as well as those measured from an independent reduction of the NIRCam grism data from the FRESCO survey \citep{Oesch_etal2023} and NIRSpec observations of JADES sources. The FRESCO spectroscopic redshifts will be discussed in Sun et al. (in prep), and the NIRSpec observations are from the JADES `DeepHST' campaign, as presented in \citep[][Bunker et al. 2023]{curtislake2023}. The literature spectroscopic redshifts come from multiple sources over the last twenty years: \citet{daddi2004, szokoly2004, mignoli2005, vanderwel2005, ravikumar2007, vanzella2007, vanzella2009, wuyts2008, balestra2010, silverman2010, xue2011, cooper2012, kurk2013, lefevre2013, trump2013, kriek2015, lefevre2015, morris2015, momcheva2016, mclure2018, pentericci2018a, pentericci2018b, wisnioski2019, garilli2021}. We also used spectroscopic redshifts from the Multi Unit Spectroscopic Explorer (MUSE) MUSE\_HUDF \citep{inami2017}, MUSE\_WIDE \citep{urrutia2019}, and MUSE\_DR2 \citep{bacon2023} surveys, and we separate those out from the rest of the literature sources. For each we only chose to compare against those spectroscopic redshifts with the most confident quality flags.

Our final spectroscopic redshift catalog consists of 263 sources from the assembled Literature catalogs, 869 from MUSE, 147 from our measurements from FRESCO data, and 150 from NIRSpec. In total there are 1429 sources (57 per square arcminute) with spectroscopic redshifts. In Figure \ref{fig:photoz_vs_specz}, we plot our estimated photometric redshifts against the assembled spectroscopic redshifts. We color the points by the source of the spectroscopic redshift, and in the legend we provide the number of sources, the number of catastrophic outliers, and the outlier fraction, defined as the number of objects with $|z_{spec} - z_{phot}| / (1 + z_{spec}) > 0.15$. At the top we provide the statistics for the overall sample. 

The comparison of the photometric redshifts to the spectroscopic redshifts is highly encouraging: we only find 68 catastrophic outliers, for an overall outlier fraction of 5\%. In many of the catastrophic outliers, the {\tt EAZY} $\chi^2$ surface has two minima and the photometry supports an incorrect photometric redshift solution. In addition, we measure the average offset between the photometric redshifts and the spectroscopic redshifts $\langle z_{spec} - z_{phot} \rangle = 0.05$, and the scatter around the relation, $\sigma_{NMAD} = 0.024$, defined as:
\begin{equation}
\sigma_{\mathrm{NMAD}} = 1.48 \times \mathrm{median}\left(\left| \frac{\delta z - \mathrm{median}(\delta z)}{1 + z_{spec}} \right| \right) 
\end{equation}
where $\delta z = z_{spec} - z_{phot}$. While the sources with high-quality spectroscopic redshifts are brighter than the bulk of the population in our sample, we don't see any evidence of systematic trends in the quality of our photometric redshifts. We do observe that the quality of the photometric redshifts is worse at $z_{spec} < 6$, but this is to be expected as the Lyman break is covered by the HST/ACS filters at these redshifts at lower sensitivities and smaller exposure times. In addition, for galaxies at $z_{spec} > 5$, Spitzer/IRAC colors revealed that galaxies have very high equivalent-width emission lines (eg. \cite{smit2014, smit2015, roberts-borsani2016, stark2016} which improve photometric redshifts significantly when properly included in templates. In Figure \ref{fig:m277w_vs_specz} we plot the observed F277W Kron magnitude against the spectroscopic redshift, and highlight the catastrophic outliers seen in Figure \ref{fig:photoz_vs_specz}. We can see that our sources with spectroscopic redshifts span a range of magnitudes down $m_{\mathrm{F277W, Kron}} = 18 - 30$, and that primarily our outliers at $z > 3$ are fainter galaxies. 

\section{Conclusions and Data Access}\label{sec:conclusions}

We have described the first NIRCam data release from JADES, which provides very deep 9-band infrared imaging of 26 arcmin$^2$ fully covering the Hubble Ultra Deep Field.  This brings the exquisite sensitivity and angular resolution of JWST to this premier deep field.  Combined with JEMS and previous HST data, there are 24 bands of space-quality optical and near-infrared imaging, revealing galaxies with great detail in their colors and morphologies.

The imaging and catalogs from this release are available at \url{https://archive.stsci.edu/hlsp/jades}. This site also includes details of the catalog contents.  Further, at \url{http://jades.idies.jhu.edu/} we provide a link to our FITSmap \citep{hausen2022a} visualization of the data, where one can pan and zoom in multiple filters,
and use overlays to present the associated catalog data.  We have found these to be extremely useful in browsing the data and in probing issues in the data reduction.

This is the first of several upcoming releases from JADES.  As described in \citet{eisenstein23r}, future work will expand the GOODS-S footprint and explore GOODS-N.  In cycle 2, the footprint of this UDF region will be observed again, roughly doubling the depth and providing for a one-year return to study the variability of the deep infrared sky, and we will conduct extensive spectroscopy in the field.

We hope that this release from JADES, as well as the companion NIRSpec release \citep{bunker23r}, will provide an important community resource for the study of the HUDF, including for the upcoming Cycle 3 proposal cycle.  The JADES data are fulfilling the dream of pushing back the redshift frontier to the first galaxies as envisaged by the early proponents of the JWST mission.  

\begin{acknowledgements}
The JADES Collaboration thanks the Instrument Development Teams and the instrument teams at the European Space Agency and the Space Telescope Science Institute for the support that made this program possible. We also thank our program coordinators at STScI for their help in planning complicated parallel observations.
\end{acknowledgements}

\begin{acknowledgements}
MR, AD, CD, EE, DJE, BDJ, BR, GR, FS, DS, ZC, LW and CNAW acknowledge support from the NIRCam Science Team contract to the University of Arizona, NAS5-02015.  DJE is further supported as a Simons Investigator.  
SAr acknowledges support from Grant PID2021-127718NB-I00 funded by the Spanish Ministry of Science and Innovation/State Agency of Research (MICIN/AEI/ 10.13039/501100011033).
AJB, AJC, JC, AS, \&  IEBW acknowledge funding from the "FirstGalaxies" Advanced Grant from the European Research Council (ERC) under the European Union’s Horizon 2020 research and innovation programme (Grant agreement No. 789056).
AJC acknowledges funding from the "FirstGalaxies" Advanced Grant from the European Research Council (ERC) under the European Union’s Horizon 2020 research and innovation programme (Grant agreement No. 789056).
\end{acknowledgements}

\begin{acknowledgements}
Funding for this research was provided by the Johns Hopkins University, Institute for Data Intensive Engineering and Science (IDIES).  
RM, WB, FDE, TJL, LS, and JW acknowledge support by the Science and Technology Facilities Council (STFC) 
and by the ERC through Advanced Grant 695671 "QUENCH". RM also acknowledges funding from a 
research professorship from the Royal Society. 
JW further acknowledges support from the Fondation MERAC.  
The research of CCW is supported by NOIRLab, which is managed by the Association of Universities for Research in Astronomy (AURA) under a cooperative agreement with the National Science Foundation.  
ALD thanks the University of Cambridge Harding Distinguished Postgraduate Scholars Programme and 
Technology Facilities Council (STFC) Center for Doctoral Training (CDT) in Data intensive science at the University of Cambridge (STFC grant number 2742605) for a PhD studentship.  
RS acknowledges support from a STFC Ernest Rutherford Fellowship (ST/S004831/1). CWo is supported by the National Science Foundation through the Graduate Research Fellowship Program funded by Grant Award No. DGE-1746060. 
\end{acknowledgements}

\begin{acknowledgements}
DP acknowledges support by the Huo Family Foundation through a P.C.\ Ho PhD Studentship.  
H{\"U} gratefully acknowledges support by the Isaac Newton Trust and by the Kavli Foundation through a Newton-Kavli Junior Fellowship. 
LW acknowledges support from the National Science Foundation Graduate Research Fellowship under Grant 
No. DGE-2137419. 
REH acknowledges acknowledges support from the National Science Foundation Graduate Research Fellowship Program under Grant No. DGE-1746060.  
SC acknowledges support by European Union’s HE ERC Starting Grant No. 101040227 -- WINGS.  
The research of KB is supported in part by the Australian Research Council Centre of Excellence for All Sky Astrophysics in 3 Dimensions (ASTRO 3D), through project number CE170100013.
\end{acknowledgements}

\begin{acknowledgements}
Processing for the JADES NIRCam data release was performed on the \emph{lux} cluster
at the University of California, Santa Cruz, funded by NSF MRI grant AST 1828315.
This work was performed using resources provided by the Cambridge Service for Data Driven Discovery (CSD3) 
operated by the University of Cambridge Research Computing Service (www.csd3.cam.ac.uk), provided by 
Dell EMC and Intel using Tier-2 funding from the Engineering and Physical Sciences Research Council 
(capital grant EP/T022159/1), and DiRAC funding from the Science and Technology Facilities Council (www.dirac.ac.uk). Support for program JWST-GO-1963 was provided in part by
NASA through a grant from the Space Telescope Science Institute,
which is operated by the Associations of Universities for Research
in Astronomy, Incorporated, under NASA contract NAS 5-26555.
\end{acknowledgements}

\bibliography{main}{}
\bibliographystyle{aasjournal}

\end{document}